\documentstyle[12pt]{article}
\input{epsf}
\newcommand{\msection}[1]{\setcounter{equation}{0}
\section[]{#1}}

\newcommand{\abs}[1]{\left\vert{#1}\right\vert}
\newcommand{\del}{\partial}
\newcommand{\delm}{\partial_\mu}
\newcommand{\deln}{\partial_\nu}
\newcommand{\nonum}{\nonumber\\}
\newlength{\mleng}
\newcommand{\mq}{\makebox[0.5\mleng]{ }}
\newcommand{\mqq}{\makebox[\mleng]{ }}
\newcommand{\mqqq}{\makebox[1.5\mleng]{ }}
\newcommand{\Lag}{{\cal L}}
\newcommand{\be}{\begin{equation}}
\newcommand{\ee}{\end{equation}}
\newcommand{\ba}{\begin{eqnarray}}
\newcommand{\ea}{\end{eqnarray}}
\newcommand{\mass}{{\cal M}}
\newcommand{\massd}{{\cal M}^{\dag}}
\newcommand{\mhatp}{\widehat{\cal M}_+}
\newcommand{\mhatm}{\widehat{\cal M}_-}
\newcommand{\gtil}{\widetilde{g}}
\newcommand{\bfq}{\makebox{\boldmath$q$}}
\newcommand{\alo}{\alpha_1}
\newcommand{\alt}{\alpha_2}
\newcommand{\alh}{\alpha_3}
\newcommand{\alpl}{\alpha_+}
\newcommand{\alm}{\alpha_-}
\newcommand{\alpp}{\alpha_p}
\newcommand{\Tr}{\mbox{Tr}}
\catcode`\@=11
\newbox\tempboxa
\newdimen\captionboxsubcount
\def\capsize#1{\captionboxsubcount=#1pt}
\newdimen\captionboxsub
\captionboxsub=\hsize \advance\captionboxsub by -\captionboxsubcount
\advance\captionboxsub by -\captionboxsubcount
\long\def\@makecaption#1#2{
 \setbox\@tempboxa\hbox{#1: #2}
 \ifdim \wd\@tempboxa >\captionboxsub
\rightskip=\captionboxsubcount \leftskip=\captionboxsubcount #1: #2
\else \hbox to\hsize{\hfil\box\@tempboxa\hfil}
 \fi}
\catcode`\@=12
\capsize{30}
\begin{document}

\thispagestyle{empty}
\begin{titlepage}

\begin{flushright}
\begin{minipage}[t]{4cm}
\begin{flushleft}
{\baselineskip = 14pt
SU-4240-613\\
hep-ph/9506473 \\
June, 1995\\
}
\end{flushleft}
\end{minipage}
\end{flushright}

\vspace{0.5cm}

\begin{center}
\Large\bf
Effects of Symmetry Breaking on the Strong and Electroweak
Interactions of the Vector Nonet
\end{center}

\vspace{0.5cm}
\vfill

\begin{center}
{\large
Masayasu {\sc Harada}\footnote{
e-mail address : {\tt mharada@npac.syr.edu}}
and
Joseph {\sc Schechter}\footnote{
e-mail address : {\tt schechter@suhep.phy.syr.edu}}
}
\\
{\it Department of Physics, Syracuse University,
Syracuse, NY 13244-1130, USA}
\end{center}

\vfill

\vspace{0.5cm}

\begin{abstract}
\baselineskip = 15pt
Starting from a chiral invariant and quark line rule conserving
Lagrangian of pseudoscalar and vector nonets we introduce first and
second order symmetry breaking as well as quark line rule violating
terms and fit the parameters, at tree level, to many strong and
electroweak processes.
A number of predictions are made.
The electroweak interactions are included in a manifestly gauge
invariant manner.
The resulting symmetry breaking pattern is discussed in detail.
Specifically, for the ``strong'' interactions,
we study all the vector meson masses and $V\rightarrow\phi\phi$
decays, including isotopic spin violations.
In the electroweak sector we study the
$\{\rho^0,\omega,\phi\}\rightarrow e^+e^-$ decays,
$\{\pi^+,K^+,K^0\}$ ``charge radii'',
$K_{l3}$ ``slope factor'' and the overall
$e^+e^-\rightarrow \pi^+\pi^-$ process.
It is hoped that the resulting model may be useful as a reasonable
description of low energy physics in the range up to about $1$ GeV.
\end{abstract}
\end{titlepage}

\setcounter{footnote}{0}

\msection{Introduction}

It seems likely that,
whether or not it can be fully derived from first principles, an
effective chiral Lagrangian will remain
as a preferred description of low energy hadronic physics.
At very low energies, near to the $\pi\pi$ threshold,
the Lagrangian can be constructed in terms of only the
pseudoscalar chiral fields.
According to the chiral perturbation
scheme\cite{Weinberg:ChPT,Gasser-Leutwyler:SU(2)-SU(3)}
it is possible to make a controlled energy expansion in which
one first keeps terms
with two derivatives or one power of the chiral
symmetry breaking quark mass.
This is the traditional chiral Lagrangian\cite{Cronin}
and is to be evaluated at tree level.
At the next stage of approximation,
single loops are computed and their divergences are cancelled by the
addition of counterterms with four derivatives or their equivalent.
Each counterterm can also make a finite contribution so there are
many arbitrary constants.
In practice it is unrealistic to go beyond four derivative order and
furthermore the finite parts of the counterterms can be largely
explained\cite{VMD}
as contributions involving vector meson terms
acting at tree level.

If one wants to extend the effective Lagrangian to describe
hadronic
physics up to about the $1$ GeV region,
it is clearly necessary to include
the additional particles -- the
vector mesons -- whose masses lie in this range.
For such Lagrangians there does not seem to be an unambiguous
controlled energy expansion.
We might imagine keeping the energy expansion around $\pi\pi$
threshold for soft interaction of the pseudoscalars.
Then, thinking of the vectors as ``heavy'' particles,
we may make an expansion around the vector meson mass shell of the
soft (chiral) interactions of the vectors.
Each of these two expansions will have a limited range of validity but
it may be possible to ``analytically continue'' between the two.
This could conceivably be extended to the entire particle spectrum.

A different way to organize the low energy Lagrangian is based on the
$1/N_{\rm c}$ approximation to QCD\cite{1/Nc}.
In this approach {\em all} \/ $\overline{q}q$ mesons
should be included
at {\em tree level} \/ at the initial stage of approximation.
The next level would contain all one loop corrections and so on.
In dealing with a limited energy range it seems
reasonable\cite{Sannino-Schechter} to
truncate the spectrum.

Regardless of whichever way is adopted for organizing the Lagrangian
it is necessary to study the effects of tree level symmetry breaking
for the vector meson nonet.
This is the main goal of the present paper.
It is, of course, an old subject.
Very recent treatments include
refs.~\cite{Schechter-Subbaraman-Weigel,%
BandoKugoYamawaki:NP,Bramon-Grau-Pancheri:symbre:95,%
Furui-Kobayashi-Nakagawa};
our treatment will follow most
closely the approach and notation of the
first of these.
There are several new features.
We shall discuss all of the vector meson masses and three point
$V\phi\phi$ coupling constants (including isospin violation effects)
as well as many of the electroweak
observables which are related to vector meson dominance.
Our enumeration of the tree level symmetry breaking terms will be more
complete than before and we shall include explicit discussion of the
relatively small Okubo-Zweig-Iizuka (OZI) rule%
\cite{OZI}
violating effects for the vectors as well as ``second order'' terms
which may be needed to fit the experimental data.
We will aim
for a more accurate and comprehensive fitting of the observable
quantities.
This will lead us to include a
number of additional diagrams for various
processes which are required by chiral invariance but contribute at
about the 10\% level and are usually neglected.
(An example is the {\em direct} \/
$\omega\rightarrow2\pi$ isospin violating vertex.)
In trying to fit the experimental numbers characterizing
the vector nonet into a reasonable theoretical pattern,
the ``sore thumb'' which sticks out is the
$K^{\ast0}$--$K^{\ast+}$ mass difference.
In ref.~\cite{Schechter-Subbaraman-Weigel} a compromise fit was
suggested to improve this prediction.
Here we will show that, even if second order terms are included, this
quantity can not be understood when other observables
are fit exactly
and when a conventional estimate is made for the photon exchange
contribution.
However,
by deriving an analog of Dashen's theorem\cite{Dashen} for
electromagnetic contributions to the vector nonet isotopic
spin violation splittings we demonstrate that there may be enough
uncertainty to save this prediction.

All the parameters of our Lagrangian will be directly fitted from
experimental data.
This includes the pure pseudoscalar field piece for which
both first
and second order OZI rule conserving terms will be treated at tree
level.
The fit for those terms essentially reproduces the standard
one\cite{Gasser-Leutwyler:SU(2)-SU(3)};
even though the latter includes the finite parts of the one loop
diagrams it doesn't contribute much owing to the conventional choice
of scale.
In general,
especially if there are many terms, it is difficult to
practically distinguish finite loop corrections from higher
order tree contributions.
We will see that either can largely explain a 30\% deviation of the
$\rho^0\rightarrow e^+e^-$ rate from its first order predicted value.

The leading symmetric terms of the effective Lagrangian are given in
section~\ref{sec: symmetric terms}.
This section also contains the introduction of the (external)
electroweak gauge bosons.
We stress that the complete Lagrangian is gauge invariant by
construction.
Section~\ref{sec: breaking terms} contains some discussion of
different approaches to counting.
We do not make a commitment to one counting
scheme or another.
However, we include enough terms to accommodate
any reasonable scheme.
We do separate the terms into these of OZI rule conserving and OZI
rule violating type.

Section~\ref{sec: parameter fitting}
discusses obtaining the parameters of the Lagrangian from experimental
values of pseudoscalar masses and decay constants and vector meson
masses and $V\rightarrow\phi\phi$ partial widths.
The needed formulae are relegated,
for the sake of readability, to Appendix~\ref{app: formula}.
The specific effects of the OZI violation terms are discussed in
section~\ref{sec: OZI violation}
(with some associated calculations described in
Appendix~\ref{app: phi pi gamma}).
It is noted that  both an OZI violating,
SU(3) conserving as well as an OZI violating,
SU(3) violating piece are needed to fit the data.
Of course, both effects are small and actually corrections due to
isospin violations are shown not to be completely negligible.

The predictions for $\Gamma(\phi\rightarrow K\overline{K})$,
$\Gamma(\omega\rightarrow \pi\pi)$,
$\Gamma(\phi\rightarrow \pi\pi)$ and
for the non electromagnetic piece of the
$K^{\ast0}$--$K^{\ast+}$ mass difference are
given in section~\ref{sec: prediction} and
Appendix~\ref{app: omega pi pi} for each value of the quark mass
ratio
$x = 2m_s / (m_u+m_d)$.
Setting $\Gamma(\omega\rightarrow\pi\pi)$ to its experimental value
leads to a standard\cite{Gasser-Leutwyler:PRep} determination of the
quark mass ratios.
It is also shown that relating
the photon exchange piece of the
$K^{\ast0}$--$K^{\ast+}$ mass difference to the $\rho^+$--$\rho^0$
mass difference permits a rather large uncertainty which may solve
the $K^{\ast0}$--$K^{\ast+}$ puzzle.
In section~\ref{sec: second order},
the effects on the fitting of including second order symmetry breaking
terms for the vectors is discussed.
It does not seem to be possible to solve the
$K^{\ast0}$--$K^{\ast+}$ puzzle in this way, however.

In section~\ref{sec: V -> ee} we discuss the $\rho^0$, $\omega$ and
$\phi$ decays into $e^+e^-$.
Even without symmetry breaking,
vector meson dominance gives a reasonable,
but not perfect,
description of these processes.
The effects of first order symmetry breaking terms unfortunately do
not perfect the descriptions.
Hence we introduce gauge invariant higher derivative photon--vector
symmetry breaking terms which enable
us to fit these decays exactly.
Once again,
a term which breaks both the OZI rule as well as SU(3) is seem to be
important.
We also remark that the $\rho^0\rightarrow e^+e^-$ production can be
improved by including the effect of the large rho width,
as pointed out a long time ago\cite{Gounaris-Sakurai}.
It is noted that this can be understood as coming from pion loop
corrections in the present framework.

Section~\ref{sec: charge radii} contains a discussion of the $\pi^+$,
$K^+$ and $K^0$ charge radii as well as the slope parameter of
$K_{e3}$ decay in our model.
The terms just mentioned which improve the $V^0\rightarrow e^+e^-$
predictions also improve the predictions for these quantities.
Both $\Gamma(\omega\rightarrow\pi\pi)$ and
$\Gamma(\rho\rightarrow e^+e^-)$ played an important role in our
analysis.
Actually these quantities are obtained from the experimental reaction
$e^+e^-\rightarrow\pi^+\pi^-$.
Thus we found it instructive to obtain the relevant Lagrangian
parameters from directly fitting our theoretical formula for this
reaction to experiment.
This is discussed in section~\ref{sec: ee -> pi pi}
and gives a feeling for the accuracy with which parameters
characterizing
a broad resonance like the $\rho$ can be extracted from
experiment.

\newpage

\msection{Effective Lagrangian
\label{sec: Effective Lagrangian}}

Before proceeding to a detailed analysis  of the symmetry breaking in
the effective chiral Lagrangian of pseudoscalars plus vectors we
shall, for the readers convenience, gather together here the
terms
which will actually be needed.
We will include what we believe to be the leading terms which
contribute to the vector meson masses and decay amplitudes.
The terms proportional to the Levi-Civita symbol
$\varepsilon_{\mu\nu\alpha\beta}$ will however not be discussed in the
present paper.

\subsection{The leading symmetric terms
\label{sec: symmetric terms}}

These obey the chiral
U(3)$_{\rm L}$$\times$U(3)$_{\rm R}$
symmetry.
They will be also considered to obey the Okubo-Zweig-Iizuka
(OZI or quark-line) rule%
\cite{OZI}
and to contain the minimum number of derivatives.
Let us start with the
U(3)$_{\rm L}$$\times$U(3)$_{\rm R}$$/$U(3)$_{\rm V}$ nonlinear
realization of chiral symmetry.
The basic quantity is a 3 $\times$ 3 matrix $U$,
which transforms as
\be
U \rightarrow U_{\rm L} U U_{\rm R}^{\dag} \ ,
\label{eq: U transformation}
\ee
where $U_{\rm L,R}\in\mbox{U(3)}_{\rm L,R}$.
This $U$ is parametrized by the pseudoscalar nonet field $\phi$ as
\be
U = \xi^2 \ , \quad\quad \xi = e^{i\phi/F_\pi} \ ,
\ee
where $F_\pi$ is a ``bare'' pion decay constant.
The vector nonet field $\rho_\mu$ is related to auxiliary linearly
transforming ``gauge fields'' $A^L_\mu$ and $A^R_\mu$
by\cite{Kaymakcalan-Schechter}
\be
A^L_\mu = \xi \rho_\mu \xi^{\dag}
  + \frac{i}{\gtil} \xi \delm \xi^{\dag} \ , \mqq
A^R_\mu = \xi^{\dag} \rho_\mu \xi
  + \frac{i}{\gtil} \xi^{\dag} \delm \xi \ ,
\label{def: A fields}
\ee
where $\gtil$ is a bare $\rho\phi\phi$ coupling.
(For an alternative approach,
which is  equivalent at tree level,
see ref.~\cite{Bando-Kugo-Yamawaki:PRep}.)
The symmetric terms may be expressed as\cite{Kaymakcalan-Schechter}
\ba
\Lag_{\rm sym}
&\equiv&
- \frac{m_v^2(1+k)}{8k} \Tr
\left[ A^L_\mu A^L_\mu + A^R_\mu A^R_\mu \right]
+ \frac{m_v^2(1-k)}{4k} \Tr
\left[ A^L_\mu U A^R_\mu U^{\dag} \right]
\nonum &&
{} - \frac{1}{4}
 \Tr\left[ F_{\mu\nu}(\rho) F_{\mu\nu}(\rho) \right] \ ,
\label{Lag: sym}
\ea
where $F_{\mu\nu}(\rho)$ is the ``gauge field strength'' of vector
mesons:
\be
F_{\mu\nu}(\rho) = \delm \rho_\nu - \deln \rho_\mu
- i \gtil [ \rho_\mu , \rho_\nu ] \ .
\ee
Notice that
$\mbox{Tr} \left[ F_{\mu\nu}(A^L) F_{\mu\nu}(A^L) \right]=$
$\mbox{Tr} \left[ F_{\mu\nu}(A^R) F_{\mu\nu}(A^R) \right]=$
$\mbox{Tr} \left[ F_{\mu\nu}(\rho) F_{\mu\nu}(\rho) \right]$
so that
(\ref{Lag: sym}) may be written directly in terms of the linearly
transforming objects $A^L_\mu$, $A^R_\mu$ and $U$.
The parameter $k$ is defined by
\be
k \equiv \left(\frac{m_v}{\gtil F_\pi}\right)^2 \ .
\label{def: k}
\ee
The choice $k=2$
(which is well-known not to follow from the requirements of chiral
symmetry)
is called the Kawarabayashi-Suzuki-Riazuddin-Fayyazuddin
(KSRF) relation\cite{KSRF}
and turns out to be close to experiment.
It is seen that the first two terms in
(\ref{Lag: sym}),
which each contain
a factor of the degenerate vector meson mass
parameter $m_v$,
break the symmetry under {\em local} \/
U(3)$_{\rm L}$$\times$U(3)$_{\rm R}$ transformations.
We may make the entire Lagrangian locally gauge invariant by
introducing a multiplet $B^{L,R}_\mu$ of external gauge fields which
transforms in the same way as $A^{L,R}_\mu$ under local
transformations.
All that is required is the simple replacement:
\be
A^{L,R}_\mu \rightarrow
A^{L,R}_\mu - \frac{h}{\gtil} B_\mu^{L,R} \ ,
\label{eq: replacement}
\ee
where $h$ is an external field coupling constant.
Of course,
the last term in eq.~(\ref{Lag: sym}) is already gauge invariant.
The electroweak gauge fields (${\cal A}_\mu$: photon,
${\cal Z}_\mu$ and ${\cal W}_\mu$: $Z$ and $W$ bosons)
are embedded as
\ba
h B^{L}_\mu &=&
e Q {\cal A}_\mu + g_2 Q_Z {\cal Z}_\mu
+ \frac{g_2}{\sqrt2}
  \left( Q_W {\cal W}_\mu^+ + Q_W^{\dag} {\cal W}_\mu^- \right)\ ,
\nonum
h B^{R}_\mu &=& e Q {\cal A}_\mu
- g_2 \frac{\sin^2\theta_W}{\cos\theta_W} Q {\cal Z}_\mu \ ,
\label{def: external}
\ea
where
\ba
&&
g_2 = - e / \sin \theta_W \ ,
\nonumber\\
&&
Q \equiv
\left(
 \begin{array}{ccc} 2/3 & & \\ & -1/3 & \\ & & -1/3 \end{array}
\right) \ ,
\mq
Q_W \equiv
\left(
 \begin{array}{ccc}
 0 & V_{ud} & V_{us} \\
 0 & 0 & 0 \\
 0 & 0 & 0
\end{array}
\right) \ ,
\nonum
&&
Q_Z \equiv \frac{1}{\cos\theta_W}
\left(
\begin{array}{ccc}
 1/2 & & \\ & -1/2 & \\ & & -1/2
\end{array}
\right)
- \frac{\sin^2\theta_W}{\cos\theta_W} \, Q \ ,
\label{def: chrages}
\ea
with $\theta_W$ being the weak angle
while
$V_{ud}$ and $V_{us}$ are appropriate
Kobayashi-Maskawa matrix elements.
Notice that (\ref{def: external})
and (\ref{def: chrages}) refer only to the three light quark degrees
of freedom and so do not include the light-heavy weak transition
currents.

It is instructive to expand to leading order the terms linear in
${\cal A}_\mu$ which are obtained after substituting
(\ref{eq: replacement})--(\ref{def: chrages})
into $\Lag_{\rm sym}$ of (\ref{Lag: sym}):
\be
e {\cal A}_\mu
\left[
  k \gtil F_\pi^2 \Tr(Q\rho_\mu) +
  i \left(1-\frac{k}{2}\right)
  \Tr\left[ Q \left( \phi
    \mathop{\partial_\mu}^{\leftrightarrow}
  \phi\right)
  \right]
\right] + \cdots \ .
\ee
We see that when $k=2$,
the direct single photon coupling to two charged pseudoscalars
vanishes.
Instead, the photon mixes with the neutral vector mesons according to
the first term and the vector mesons then couple to two pseudoscalars.
Hence, to the extent that $k$ is experimentally equal to $2$,
a natural notion of vector meson dominance is automatic in this
framework.
By construction, the model is gauge invariant.
Furthermore a transformation to the zero mass physical photon basis
may be seen \cite{Bando-Kugo-Yamawaki:PRep,Schechter:PRD34}
to give the identical results.

In writing (\ref{Lag: sym}) we considered only terms at most bilinear
in $A^L_\mu$ and $A^R_\mu$.
Other symmetric terms with pieces such as
$\Tr\left[F_{\mu\nu}(A^L)A^L_\mu A^L_\mu\right]$,
$\Tr\left[A^L_\mu A^L_\mu A^L_\nu A^L_\nu\right]$ etc.
are formally suppressed in the $1/N_c$ expansion.

\subsection{Symmetry breaking terms
\label{sec: breaking terms}}

We next consider chiral U(3)$_{\rm L}$$\times$U(3)$_{\rm R}$
breaking
terms which reflect the presence of non-zero quark masses in the
fundamental QCD Lagrangian.
We shall also consider terms which are not single traces in the
SU(3) flavor space and thereby violate the OZI rule.

The fundamental QCD Lagrangian contains the
quark mass term
$-\widehat{m} \overline{q} \mass q$, where
$\widehat{m}\equiv(m_u+m_d)/2$, and $\mass$ is the dimensionless
matrix:
\be
\mass = \left(
\begin{array}{ccc}
1+y & & \\
& 1-y & \\
& & x
\end{array}
\right) \ .
\ee
Here
$x$ and $y$ are the quark mass ratios:
\be
x = \frac{m_s}{\widehat{m}} \ ,  \mq
y = - \frac{1}{2} \left( \frac{m_d-m_u}{\widehat{m}} \right) \ .
\ee
Another quantity of interest is
\be
R = \frac{m_s - \widehat{m}}{m_d-m_u} = \frac{1-x}{2y} \ .
\ee
We shall treat $\mass$ as a ``spurion'' which transforms like $U$ in
eq.~(\ref{eq: U transformation}).
Then the symmetry breaking
terms are {\em formally} \/ chiral invariant.

The question of which symmetry breaking terms to include is,
of course, a crucial one.
First,
consider the case when the vectors are absent.
Then, the conventional approach is based on the
chiral perturbation (ChPT)
scheme\cite{Weinberg:ChPT,Gasser-Leutwyler:SU(2)-SU(3)}
in which one makes an expansion around threshold in such a way that
two derivatives count the same as one appearance of $\mass$.
Then it seems that an adequate treatment of pseudoscalar
masses and decay constants can be made by going to 4-derivative
(or $\mass^2$) order and choosing a scale
in such a way that loop diagrams give negligible contributions.
It is necessary to give special consideration to the very important
OZI rule violation (U(1) problem) in the pseudoscalar multiplet
--- this was discussed in the present framework in sect.~IV of
ref.~\cite{Schechter-Subbaraman-Weigel}
and will not be repeated here.
Apart from this,
there are three OZI rule conserving,
pseudoscalar-type symmetry breakers to order $\mass^2$.
Specifically, at order ${\cal M}$ there is one which is linear in the
quark mass matrix without any derivatives, while at next order there
is a quadratic in ${\cal M}$ piece with no derivatives and a linear
in ${\cal M}$ piece with two derivatives.
These are,
as we will again see here,
sufficient to reproduce the usual chiral perturbation theory fit to
the pseudoscalar symmetry breaking parameters.

Now let us add the vectors.
There is no immediately obvious counting scheme that will enable us to
describe physics well away from $\pi\pi$ threshold.
If there were it would amount to a practical solution of low energy
QCD.
However there are certainly useful clues to this problem.
In the large $N_c$ approximation of QCD
the effective Lagrangian contains all nonet mesons and the leading
approximation consists of keeping just the tree diagrams.
In the low energy region
this amounts to adding the vectors with presumably minimal
derivative terms.
The $N_c$ expansion approach tends to replace the higher derivative
pseudoscalar terms with the vectors.
As far as symmetry breaking terms are concerned,
it seems reasonable to emulate the pseudoscalar case and consider OZI
rule conserving terms linear in ${\cal M}$ with no derivatives,
quadratic in ${\cal M}$ with no derivatives and linear in ${\cal M}$
with two derivatives.
We shall see that there are considerably more than three terms of
these types.
Another
approximation method is useful if we are mainly interested in soft
pionic interactions of ``on-shell'' vector mesons.
Then we may imagine the vectors to be heavy and make a chiral
perturbation expansion around that point.
For proper counting we should take the ``heavy'' vector to be
moving with fixed 4-velocity $V_\mu$ and
count the momentum as that of the ``fluctuation'' field
$\rho'_\mu = e^{-im_\rho V\cdot x} \rho_\mu$.
The resultant counting is similar to counting derivatives.
Of course, considering the vectors as ``heavy'' is debatable.
In any event, the terms we use will give rise to the heavy field ones
in the appropriate limit.%
\footnote{
If $h=h'e^{iMV\cdot x}$ is a generic heavy boson field,
the free Lagrangian
$-\partial_\mu h \partial_\mu \overline{h}
- M^2 h \overline{h}$
may be rewritten as
$- i M V_\mu h'
{\displaystyle \mathop{\partial_\mu}^{\leftrightarrow}}
\overline{h'}
+ \cdots$.
This is the leading term of the ``heavy'' meson Lagrangian,
which is to be expressed in terms of the field $h'$.
A non derivative symmetry breaker (like the $\alpha_{\pm}$ terms of
table~~\ref{tab: new para}) similarly gives rise to a term like
$m_q M h' {\cal M} \overline{h'}$
which is counted of order quark mass $m_q$ or momentum $k^2$.
A derivative type symmetry breaker (like the $\gamma'$ term
in eq.~(\ref{Lag: first})) is rewritten as
$\frac{m_q}{M} \partial_\mu h {\cal M} \partial_\mu \overline{h}
= - m_q M h' {\cal M} \overline{h'}
+ i m_q V_\mu h' {\cal M}
{\displaystyle \mathop{\partial_\mu}^{\leftrightarrow}}
\overline{h'}
+ \cdots$.
The first term on the right hand side has the same structure as the
non derivative symmetry breaker.
The second term on the right hand side has the same structure as an
order $k^3$ term in the heavy meson counting.
It is however formally suppressed by a factor $1/M$ compared to the
leading terms.
In practice, since the vectors are not very heavy, it plays the role
of such a term.
\label{foot: heavy}
}

Our initial model will contain the OZI rule conserving vector meson
terms linear in ${\cal M}$,
both with no derivatives and two derivatives.
The vector meson terms of order ${\cal M}^2$ will be discussed
separately.
We only consider vector meson terms at most bilinear in the vector
fields.
Thus we write:
\ba
\Lag_{\rm SB}
&=& \alo \Tr
\left[
 \massd A^L_\mu U A^R_\mu + \mass A^R_\mu U^{\dag} A^L_\mu
\right]
\nonum
&& + \alt \Tr
\left[
 U^{\dag} \mass U^{\dag} A^L_\mu U A^R_\mu
 + U \massd U A^R_\mu U^{\dag} A^L_\mu
\right]
\nonum
&& + \frac{\alh}{2} \Tr
\left[
 \left(\mass U^{\dag} + U \massd \right) A^L_\mu A^L_\mu
 + \left(\massd U + U^{\dag} \mass \right) A^R_\mu A^R_\mu
\right]
\nonum
&& +
\gamma' \Tr
\left[
 \massd F^L_{\mu\nu} U F^R_{\mu\nu}
 + \mass F^R_{\mu\nu} U^{\dag} F^L_{\mu\nu}
\right]
\nonum
&& + \delta'\Tr
\left[ \mass U^{\dag} + \massd U \right]
\nonum
&& + {\lambda'}^2 \Tr
\left[
  \mass U^{\dag} \mass U^{\dag} +
  \massd U \massd U - 2 \massd \mass
\right] \ .
\label{Lag: first}
\ea
The $\delta'$ term is the standard one for the pseudoscalar masses
and the $\lambda'^2$ gives $\mass^2$ corrections to it.
The third purely pseudoscalar symmetry breaker of the type
$\mass\partial^2$ is, noting eq.~(\ref{def: A fields}),
hidden in the $\alpha_i$ terms
[see Appendix~\ref{app: rewriting}].
There are two remaining independent non-derivative vector meson
symmetry breaking combinations among the $\alpha_i$ terms
corresponding to mass and $\rho\phi\phi$ coupling constant splittings.
Previous fits\cite{Schechter-Subbaraman-Weigel,%
BandoKugoYamawaki:NP,%
Bramon-Grau-Pancheri:symbre:95,Furui-Kobayashi-Nakagawa}
of vector meson properties
have only included a single linear
combination of these.
The new term noticeably improves the overall fit.
Finally, the $\gamma'$ term represents symmetry breaking in the vector
meson kinetic terms.
Another term of this type,
obtained by replacing $\mass\rightarrow U\massd U^{\dag}$,
is not independent
as
may be seen with the help of the relations,
$F^L_{\mu\nu}= \xi F_{\mu\nu}(\rho)\xi^{\dag}$ and
$F^R_{\mu\nu}= \xi^{\dag} F_{\mu\nu}(\rho)\xi$.

As previously noted,
chiral counting and the inclusion of pion loops can be consistently
done for pseudoscalars
very close to the $\pi\pi$ threshold and also for
very soft vector pion
interactions
wherein the vector remains close to its mass shell.
However we shall initially restrict
ourselves to tree diagrams
(as in the $1/N_c$ expansion)
while using the effective Lagrangian in the
extended low energy region up to about 1 GeV.
Although we will not use it for our first fit,
we give here
the non-derivative $\mass^2$ terms for the vectors:
\ba
\Lag_{\rm \mu}
&=&
\mu_1 \Tr \left[ A^L_\mu \mass A^R_\mu \massd \right]
\nonum
&& +
\frac{\mu_2}{2} \Tr \left[
  A^L_\mu U \massd U A^R_\mu \massd +
  A^R_\mu U^{\dag} \mass U^{\dag} A^R_\mu \mass
\right]
\nonum
&& +
\mu_3 \Tr \left[
  A^L_\mu U \massd U A^R_\mu U^{\dag} \mass U^{\dag}
\right]
\nonum
&& +
\frac{\mu_4}{2} \Tr \left[
  A^L_\mu U \massd A^L_\mu \mass U^{\dag} +
  A^R_\mu U^{\dag} \mass A^R_\mu \massd U
\right]
\nonum
&& +
\frac{\mu_5}{4} \Tr \Bigl[
  A^L_\mu U \massd A^L_\mu U \massd
  + A^L_\mu \mass U^{\dag} A^L_\mu \mass U^{\dag}
\nonum && \mq
  + A^R_\mu U^{\dag} \mass A^R_\mu U^{\dag} \mass
  + A^R_\mu \massd U A^R_\mu \massd U
\Bigr] \ .
\label{Lag: second}
\ea

We would also like to include terms which describe the small OZI rule
violation in the vector nonet
($m_\omega-m_\rho \simeq 12$MeV compared to
$m_{K^{\ast}}-m_\rho \simeq 124$MeV).
The terms which violate this rule but are
nevertheless chiral symmetric
take the form:
\be
\Lag_{\nu 1} \equiv
\nu_1 \left[(\Tr[A^L_\mu])^2+(\Tr[A^R_\mu])^2\right] +
\nu_2 \Tr[A^L_\mu] \Tr[A^R_\mu] \ .
\label{Lag: nu 1}
\ee

We also give the leading OZI violating terms\footnote{
There exist two other terms
which violate the OZI rule and break SU(3) symmetry:
$\mbox{Tr}\, \bigl[ U \massd + U^{\dag} \mass \bigr]
\mbox{Tr}\, \bigl[ A^L_\mu A^L_\mu + A^R_\mu A^R_\mu\bigr]$,
$\mbox{Tr}\, \bigl[ U \massd + U^{\dag} \mass \bigr]
\mbox{Tr}\, \bigl[ A^L_\mu U A^R_\mu U^{\dag} \bigr]$.
However,
for the physical quantities studied in this paper,
(vector meson masses, $\rho\pi\pi$ coupling and so on),
these effects are absorbed into the coefficients
of the symmetric terms in eq.~(\ref{Lag: sym}).
Hence we do not include such terms explicitly.
}
proportional to
$\mass$:
\ba
\lefteqn{
\Lag_{\nu 2} \equiv
\nu_3 \Biggl\{
\Tr[A^L_\mu]
\Tr[A^L_\mu U \massd + A^L_\mu \mass U^{\dag}]
+
\Tr[A^R_\mu]
\Tr[A^R_\mu U^{\dag} \mass + A^R_\mu \massd U]
\Biggr\}
}
\nonum
&& {}+
\nu_4 \Biggl\{
\Tr[A^L_\mu]
\Tr[A^R_\mu U^{\dag} \mass + A^R_\mu \massd U]
+
\Tr[A^R_\mu]
\Tr[A^L_\mu U \massd + A^L_\mu \mass U^{\dag}]
\Biggr\} \ .
\label{Lag: nu 2}
\ea
One might expect, at first,
the terms in $\Lag_{\nu2}$ to be very much suppressed compared to
those in $\Lag_{\nu1}$.
We will see later whether this, in fact, holds.

Linear combinations of the symmetry breaking parameters introduced in
this section
which naturally appear in the computation of physical
quantities are listed in table~\ref{tab: new para}.
\begin{table}[htbp]
\begin{center}
\begin{tabular}{|l|l|}
\hline
$\alpl$ & $\alo + \alt + \alh$ \\
$\alm$ & $\alo - \alt$ \\
$\alpp$ & $\alo + \alt - \alh$ \\
\hline
$\mu_a$    & $\mu_1 + \mu_2 + \mu_3 + \mu_4 + \mu_5$ \\
$\mu_b$    & $\mu_1 - \mu_2 + \mu_3 + \mu_4 - \mu_5$ \\
$\mu_c$    & $\mu_1 + \mu_2 + \mu_3 - \mu_4 - \mu_5$ \\
$\mu_d$    & $\mu_1 - \mu_2 + \mu_3 - \mu_4 + \mu_5$ \\
$\mu_e$    & $\mu_1 - \mu_3$ \\
\hline
$\nu_a$    & $2\nu_1 + \nu_2$ \\
$\nu_b$    & $4\nu_3+4\nu_4$ \\
$\nu_c$    & $2\nu_1 - \nu_2$ \\
$\nu_d$    & $4\nu_3-4\nu_4$ \\
\hline
\end{tabular}
\end{center}
\caption[]{
Convenient parameter combinations.
\label{tab: new para}}
\end{table}
It should be stressed that the substitution
(\ref{eq: replacement}) in all the terms above simply accomplishes the
task of introducing electroweak interactions in a gauge invariant way.

\newpage

\msection{Observable Quantities of the Model
\label{sec: fit}}

In this section
we discuss the physical quantities computed
from the Lagrangian
\be
\Lag = \Lag_{\rm sym} + \Lag_{\rm SB} + \Lag_{\nu1} + \Lag_{\nu2} \ .
\label{eq: Lagrangian}
\ee

Altogether we consider 13 {\em a priori} \/ unknown parameters.
At the level of the symmetric piece $\Lag_{\rm sym}$,
there are only three quantities:
$m_v$, $\gtil$ and the ``bare'' pion decay constant $F_\pi$.
Adding symmetry breaking brings in the two quark mass ratios
$x$ and $y$.
Our analysis yields, as a byproduct, an alternative extraction of
these fundamental quantities from experiment.
There are three OZI rule conserving symmetry breaking coefficients
($\delta'$, $\lambda'^2$ and $\alpha_p$)
associated with the pure pseudoscalar sector and three more OZI rule
conserving but symmetry breaking coefficients
($\alpha_+$, $\alpha_-$ and $\gamma'$)
resulting from the addition of vectors.
Two coefficients ($\nu_a$ and $\nu_b$) describing OZI rule violation
for the vector multiplet bring the total to thirteen.

The results of computing the needed physical quantities
and related discussions
are presented
in Appendix~\ref{app: formula}.

\subsection{Parameter fitting
\label{sec: parameter fitting}}

The introduction of symmetry breaking terms requires us to renormalize
the various fields.
First, we consider the
isospin symmetric limit by setting $y=0$ in
the breaking term $\mass$.
The $\alpp$-term
and $\gamma'$-term
give contributions to the kinetic terms
of the pseudoscalar mesons and the vector mesons,
respectively.
Then taking typical examples:
\ba
&&
\pi^+ \equiv Z_\pi \phi_{12} \ , \
K^+ \equiv Z_K \phi_{13} \ ,
\nonum
&&
\rho_\mu^+ = Z_{\rho} \rho_{12\mu} \ ,  \
K_\mu^{\ast+} = Z_{K^{\ast}} \rho_{13\mu} \ , \
\omega_\mu = Z_{\omega}
  (\rho_{11\mu}+\rho_{22\mu}) / \sqrt{2} \ ,  \
\phi_\mu = Z_{\phi} \rho_{33\mu} \ . \
\label{eq: field renorm}
\ea
The explicit forms of the
normalization constants are
shown in eqs.~(\ref{def: Zpi}) and (\ref{def: Zrho}).
The renormalizations of the pseudoscalar fields imply
that physical pion and kaon decay constants
$F_{\pi p}$ and $F_{Kp}$ are also renormalized as
\be
F_{\pi p} = Z_\pi F_\pi\ , \mq
F_{Kp} = Z_K F_\pi \ .
\label{def: decay constants}
\ee

We will determine the parameters of the model using the most well
known quantities namely the particle masses, $\rho$
and $K^{\ast}$ decay
widths\footnote{
Actually as discussed on p.~1456 of ref.~\cite{ParticleDataGroup:94},
the precise values of mass, width etc. for very broad resonances
depend to some extent on the method of parameterization.
See also section~\ref{sec: ee -> pi pi} of the present paper.
}
and the decay constants of (\ref{def: decay constants}).
At first we neglect OZI rule violations;
it will be discussed separately in the next subsection.
The inputs are
displayed in table~\ref{tab: physical input}.
All but one of them will
(for our present purpose)
have negligible errors.
However, the non-electromagnetic piece of the $K^0$--$K^+$ mass
difference, contains the error
shown\cite{Gasser-Leutwyler:PRep}
associated with the theoretical estimation of the photon exchange
piece.
\begin{table}[htbp]
\begin{center}
vector meson masses and partial widths (MeV)\\
\begin{tabular}{|cccc|cc|}
\hline\hline
$m_\rho$ & $m_\omega$ & $m_{k^{\ast}}$ & $m_\phi$
& $\Gamma(\rho\rightarrow\pi\pi)$ &
$\Gamma(K^{\ast}\rightarrow K\pi)$ \\
\hline
769.9 & 781.9 & 893.8 & 1019.4
& 151.2 & 49.8 \\
\hline\hline
\end{tabular}\\
\vspace{0.2cm}
pseudoscalar meson masses and decay constants(MeV)\\
\begin{tabular}{|cc|c|cc|}
\hline\hline
$m_\pi$ & $m_K$ &
$\left[\Delta m_K\right]_{\rm nonEM}$
& $F_{\pi p}$ & $F_{Kp}$ \\
\hline
137.3 & 495.7 & $5.27\pm0.30$ &
130.7 & 159.8 \\
\hline\hline
\end{tabular}
\end{center}
\caption[]{The physical inputs used for fitting.
The values are listed in the Particle Data Group
\cite{ParticleDataGroup:94}.
\label{tab: physical input}
}
\end{table}

The pion and kaon masses
are obtained by expanding the
$\delta'$ and $\lambda'^2$ terms:
\ba
 m_{\pi}^2 &=& \frac{8}{F_{\pi p}^2}
 \left(\delta' + 4 {\lambda'}^2 \right) \ ,
\nonum
 m_K^2 &=& \frac{4}{F_{K p}^2}
 \left[ (1+x) \delta' + 2 (1+x)^2 {\lambda'}^2 \right]
\ .
\ea
The contributions to the
vector meson masses  (see eq.~(\ref{eq: vector masses}))
and the $\rho\phi\phi$ coupling constants
(see eqs.~(\ref{def: V PP})--(\ref{def: direct phi pipi}))
come from the $\alpha_+$, $\alpha_-$ and $\nu$ terms.
The theoretical expression for
$\left[\Delta m_K\right]_{\rm nonEM}$
is finally given in
eq.~(\ref{eq: K0 - K+ mass diff}).

First, we determine
the values of all parameters for various fixed values of $x$.
Using the physical values of the two decay constants
$F_{Kp}$ and $F_{\pi p}$,
we determine $\alpha_p/\gtil^2$
and $F_\pi^2$.
The parameters
$m_v$, $\alpha_+$ and $\gamma'$
are determined from the vector meson masses
$m_\rho$, $m_{K^{\ast}}$ and $m_\phi$.
Next, the parameters $\alpha_-$ and $\gtil$ are determined from the
widths
$\Gamma(\rho\rightarrow\pi\pi)$  and
$\Gamma(K^{\ast}\rightarrow K\pi)$.
The isospin breaking parameter $y$ is finally
determined from the non-electromagnetic part of the
$K^0$--$K^+$ mass difference.
We list the values of parameters in table~\ref{tab: parameters}
for each choice of $x$.
(We list also the quark mass ratio $R=(1-x)/(2y)$.)
\begin{table}[htbp]
\footnotesize
\begin{center}
\begin{tabular}{|c||c|c||c|c|c|c|c|c|c|c|c|}
\hline
 & &  & $\gamma'$  & $\alpha_+$$^{\dag}$
 & $\alpha_-$$^{\dag}$  & $\alpha_p$$^{\dag}$
 & $\delta'$$^{\dag\dag}$  & $\abs{\lambda'}$$^{\dag}$
 & $m_v$
 &  & $F_\pi$ \\
 $x$ & $y$ & $R$  & $\times10^{-3}$  & $\times10^{-3}$ & $\times10^{-3}$
 & $\times10^{-3}$  & $\times10^{-3}$
 & $\times10^{-3}$
 & (GeV) & $\widetilde{g}$  & (GeV) \\
\hline\hline
$14$ & $-0.12$ & $54.1$ & $0.419
$ & $-7.23$ & $-0.684$ & $5.27
$ & $0.0304$ & $1.57
$ & $0.75$ & $4.03
$ & $0.126$ \\
$17$ & $-0.155$ & $51.5$ & $0.345
$ & $-5.87$ & $-0.542$ & $4.3
$ & $0.0344$ & $1.21
$ & $0.753$ & $4.04
$ & $0.127$ \\
$20.5$ & $-0.202$ & $48.2$ & $0.29
$ & $-4.81$ & $-0.432$ & $3.54
$ & $0.0369$ & $0.916
$ & $0.756$ & $4.04
$ & $0.127$ \\
$23$ & $-0.241$ & $45.6$ & $0.264
$ & $-4.25$ & $-0.375$ & $3.15
$ & $0.038$ & $0.755
$ & $0.758$ & $4.05
$ & $0.128$ \\
$26$ & $-0.295$ & $42.4$ & $0.241
$ & $-3.73$ & $-0.32$ & $2.77
$ & $0.0388$ & $0.597
$ & $0.759$ & $4.05
$ & $0.128$ \\
$29$ & $-0.358$ & $39.2$ & $0.225
$ & $-3.32$ & $-0.276$ & $2.48
$ & $0.0394$ & $0.464
$ & $0.761$ & $4.05
$ & $0.128$ \\
$32$ & $-0.433$ & $35.8$ & $0.214
$ & $-2.98$ & $-0.24$ & $2.24
$ & $0.0398$ & $0.343
$ & $0.761$ & $4.05
$ & $0.129$ \\
\hline
\end{tabular}
\end{center}
\caption[]{
The values of parameters determined from the experimental data.
The unit of the quantities indicated by $^{\dag}$ ($^{\dag\dag}$)
is (GeV)$^2$ ((GeV)$^4$).
}
\label{tab: parameters}
\end{table}
It should be remarked
that the numerical values in table~\ref{tab: parameters}
already include the corrections obtained
when fitting in the presence of OZI rule violation.

In our model the following generalization of the Gell-Mann--Okubo
mass formula holds:
\be
Z_\phi^2 m_\phi^2 =
2 Z_{K^{\ast}}^2 m_{K^\ast}^2 - Z_\rho^2 m_\rho^2 \ ,
\label{mass rel vector 2}
\ee
Actually, in our fit,
$Z_\phi$, $Z_{K^{\ast}}$ and $Z_\rho$ do not differ too much from
unity.
(Noting that $Z_{K^{\ast}}^2 = (Z_\phi^2 + Z_\rho^2)/2$,
we see that the ratio $Z_\phi/Z_\rho$ is independent of parameter
choice.)
In an earlier fit\cite{Schechter-Subbaraman-Weigel}
larger deviations of $Z_\phi$ and $Z_{K^{\ast}}$ from unity
(corresponding to larger choices for $x$)
were considered in an attempt to explain the puzzling value of the
non-electromagnetic piece of the $K^{\ast0}$--$K^{\ast+}$ mass
difference.
Here we will show that the electromagnetic contribution to
$K^{\ast0}$--$K^{\ast+}$ may be more uncertain than previously
thought.
(see section~\ref{sec: prediction})

As one might expect, the parameters $m_v$, $\gtil$ and $F_\pi$ of the
symmetric Lagrangian turn out not to depend much on $x$.
We will find a best value of $x$ by later examining predictions of the
model.
Notice that the $\alpha_-$ coefficient turns out to be only about
$1/10$ of $\alpha_+$.
$\alpha_+$ contributes to both mass and coupling constant
splittings while $\alpha_-$ contributes only to coupling constant
splittings.
This shows that both types of splittings are approximately controlled
by the parameter $\alpha_+$.
This may be understood by noting from
eqs.~(\ref{eq: vector masses}) and (\ref{eq: V phi phi coupling})
that $\alpha_-\neq0$ corresponds to a deviation
from the formula
$g_{\rho\pi\pi}=m_\rho^2/(\gtil F_{\pi p}^2)$.

\subsection{OZI rule violation for the vector nonet
\label{sec: OZI violation}}

\begin{figure}[htbp]
\begin{center}
\ \epsfbox{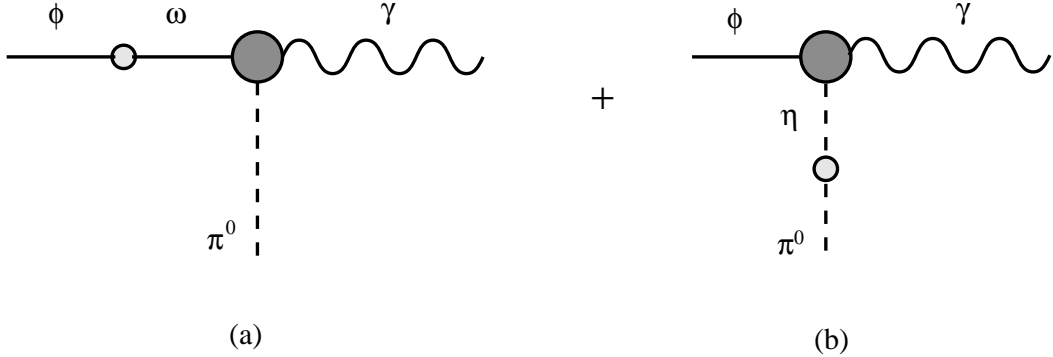}
\end{center}
\caption[]{Feynman diagrams describing the
contributions to
$\phi\rightarrow\pi^0\gamma$ decay from
(a)~$\phi$--$\omega$ mixing and (b)~$\pi^0$--$\eta$ mixing.
}
\label{fig: phi -> pi gamma}
\end{figure}

Historically the OZI rule was discovered in trying to understand the
vector meson nonet, which emphasizes that
these violating effects are small.
The $\nu_a$ term in eq.~(\ref{Lag: nu term}) yields SU(3) symmetric OZI
rule violation while the $\nu_b$ term yields SU(3) non-symmetric OZI
rule violation.
We will fit these two parameters from $m_\omega-m_\rho$ and from an
{\em experimentally determined} \/ $\omega\phi$ mixing angle.
The latter is defined from
\be
\left(
\begin{array}{c}
 \omega_\mu \\
 \phi_\mu \\
\end{array}
\right)
=
\left(
\begin{array}{cc}
 \cos \theta_{\phi\omega} & \sin \theta_{\phi\omega} \\
 - \sin \theta_{\phi\omega} & \cos \theta_{\phi\omega} \\
\end{array}
\right)
\left(
\begin{array}{c}
 \omega_{p\mu} \\
 \phi_{p\mu} \\
\end{array}
\right) \ ,
\label{eq: phi omega fields}
\ee
where the subscript $p$ denotes the physical field.
Furthermore our convention (see eq.~(\ref{eq: field renorm}))
sets
$\omega_\mu$ and $\phi_\mu$ to be the ``ideally mixed'' fields.
Hence the mixing angle $\theta_{\phi\omega}$ will be very small.
It can be seen from eq.~(\ref{eq: vector masses}) that
$m_\omega-m_\rho$ determines the combination $(\nu_a+\nu_b)$
while eq.~(\ref{eq: phi omega mix}) shows that the mixing angle will
determine the combination $2\nu_a+(1+x)\nu_b$.
First it is interesting to see what $\theta_{\phi\omega}$ would be if
$\nu_b$ were absent.
Then we may calculate
\be
\theta_{\phi\omega} \simeq
\frac{\Pi_{\phi\omega}}{m_\phi^2 - m_\omega^2}
=
\frac{m_\omega^2-m_\rho^2}{m_\phi^2-m_\omega^2}
\sqrt{\frac{m_\phi^2-m_{K^{\ast}}^2}%
{2m_{K^{\ast}}^2 - 3m_\rho^2 + m_\omega^2}}
\simeq 0.0325 \ ,
\label{eq: phi omega angle}
\ee
where the relation
\be
\left(\frac{Z_\phi}{Z_\rho}\right)^2
=
\frac{m_{K^\ast}^2 - ( 3m_\rho^2 - m_\omega^2)/2}
{ m_\phi^2 - m_{K^\ast}^2}
\label{mass rel vector 3}
\ee
was used.
While the numerical value of $\theta_{\phi\omega}$ in
eq.~(\ref{eq: phi omega angle}) has the right order of magnitude,
it turns out to be only about 60\% of the value needed to explain the
branching ratio
$\Gamma(\phi\rightarrow\pi^0\gamma)/%
\Gamma(\omega\rightarrow\pi^0\gamma)$ using a model based on the
mixing approximation.\footnote{
It seems plausible to neglect the effects
of a direct OZI violating term of the form
$\varepsilon_{\mu\nu\alpha\beta}$$
\Tr\left(\partial_\mu\rho_\nu\right)
\Tr\left(\partial_\alpha\rho_\beta\phi\right)$,
which could lead to
$\phi\rightarrow\pi^0\rho\rightarrow\pi^0\gamma$.
}
Inclusion of the $\nu_b$ term can lead to experimental agreement.
The relevant Feynman diagram is shown in
fig.~\ref{fig: phi -> pi gamma},
along with an iso-spin violating one which gives a 10\% correction.
It is amusing that iso-spin violations may be non-negligible;
this is due to the smallness of OZI rule violation.
The calculation is discussed in Appendix~\ref{app: phi pi gamma}.
The iso-spin violating correction depends on $x$ and this leads to a
small $x$ dependence for the ``experimentally obtained'' value of
$\theta_{\phi\omega}$.
As $x$ varies from $14$ to $32$,
$\theta_{\phi\omega}$ varies from $0.056$ to $0.052$ radians.
The choice $x=20.5$ leads to the OZI rule violating parameters
\ba
\nu_a &=& -4.29 \times 10^{-3} \mbox{GeV}^2 \ , \nonumber\\
\nu_b &=& -0.357 \times 10^{-3} \mbox{GeV}^2 \ .
\ea
While $\nu_b$ makes a small contribution to $m_\omega-m_\rho$,
it actually (because it gets multiplied by $x$)
makes a very substantial contribution to $\theta_{\phi\omega}$.
It thus appears that the product (OZI violation) $\times$
(SU(3) symmetry breaking) is not suppressed for the vector nonet as
one might initially expect.

Among the parameters shown in table~\ref{tab: parameters},
only $\gamma'$ and $\alpha_-$
have a non-negligible dependence on OZI violation.
These dependences are illustrated,
for $x=20.5$, in table~\ref{tab: nu dependence}.
Also shown are the effects on some physical observables which will be
discussed later.
\begin{table}[htbp]
\footnotesize
\begin{center}
\begin{tabular}{|c||c|c|c||c|c|c|}
\hline
 & $\gamma'$ & $\alpha_-$  & $Z_\phi$
 & $\left[\mbox{Br}(\phi)\right]_{\pi\pi}$
 & $\Gamma(\omega\rightarrow e^+e^-)$
 & $\left\langle r_{K^0}^2 \right\rangle$
 \\
 & $\times10^{-3}$ & $\times10^{-3}$(GeV)$^2$ &
 & $\times10^{-4}$ & (KeV) & (fm)$^2$ \\
\hline\hline
(a)& $0.905$ & $0.0964$ & $0.923$ & $0$     & $0.583$ & $-0.0150$\\
(b)& $0.660$ & $-0.112$ & $0.944$ & $0.116$ & $0.638$ & $-0.00877$\\
(c)& $0.290$ & $-0.432$ & $0.976$ & $0.869$ & $0.682$ & $-0.00141$\\
\hline
\end{tabular}
\end{center}
\caption[]{
The dependence of the parameters and physical observables
on the OZI violating parameters $\nu_a$ and $\nu_b$.
We show the values for
(a)~$\nu_a=\nu_b=0$; (b)~$\nu_a\neq0$ and $\nu_b=0$;
(c)~$\nu_a\neq0$ and $\nu_b\neq0$
with $x=20.5$ fixed.
}
\label{tab: nu dependence}.
\end{table}

\subsection{Predictions
\label{sec: prediction}}

\begin{figure}[htbp]
\begin{center}
\ \epsfbox{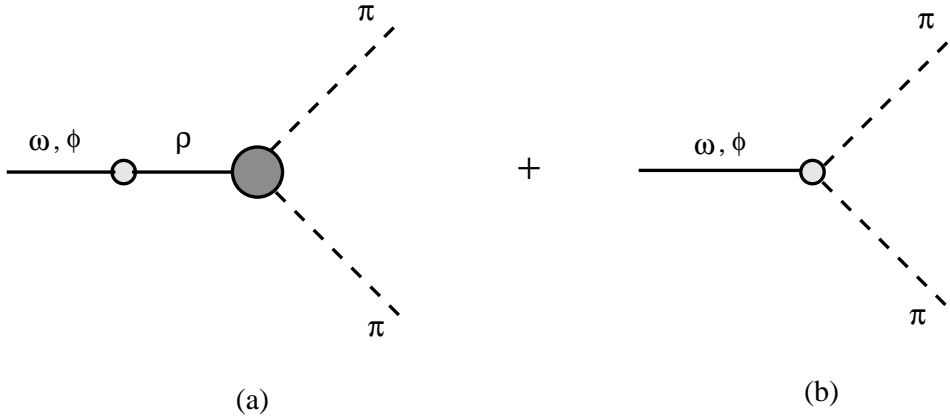}
\end{center}
\caption[]{Feynman diagrams describing the contributions to
$\omega (\phi) \rightarrow \pi\pi$ decay
from (a) $\rho^0$--$\omega$ ($\phi$) mixing
and (b) direct $\omega\pi\pi$ ($\phi\pi\pi$) vertex.
}\label{fig: w -> pi pi}
\end{figure}

So far we used up most of the vector meson masses and widths just to
determine the coefficients of the Lagrangian (\ref{eq: Lagrangian})
for any $x$.
However, there are several physical masses and widths left over
which
we can predict and compare with experiment.
This will also enable us to choose the value of $x$.
The leftover masses are the non-electromagnetic parts of
$\Delta m_{K^{\ast}} \equiv m(K^{\ast0}) - m(K^{\ast+})$
and the $\rho^0$--$\omega$ and $\rho^0$--$\phi$ transition masses.
These actually contribute to the $\omega\rightarrow2\pi$
and $\phi\rightarrow2\pi$ decays according to the diagram of
fig.~\ref{fig: w -> pi pi}(a).
In the present model there are also direct $\omega\rightarrow2\pi$
and $\phi\rightarrow2\pi$ vertices as shown in
fig.~\ref{fig: w -> pi pi}(b).
Another predicted three point vertex describes
$\Gamma(\phi\rightarrow K\overline{K})$;
in this case there is also a correction corresponding to
$\phi\rightarrow \omega\rightarrow K\overline{K}$.

We give the predictions for
$\Gamma(\phi\rightarrow K\bar{K})$,
$\Gamma(\omega\rightarrow\pi\pi)\,
/\,\Gamma(\rho\rightarrow\pi\pi)$
and the non-electromagnetic part of the
$K^{\ast0}$--$K^{\ast+}$ mass difference $\Delta m_{K^{\ast}}$
in table~\ref{tab: predictions 1}.
We include the OZI rule violating $\nu$ terms,
which generate the $\rho$--$\phi$ mixing and the direct $\phi\pi\pi$
coupling, in our fit.
Also shown is the predicted
branching ratio $\Gamma(\phi\rightarrow\pi\pi)/
\Gamma(\phi\rightarrow\mbox{\rm all})$.
\begin{table}[htbp]
\begin{center}
\small
\begin{tabular}{|c||c||c|c|c||c||c|}
\hline
 & $\Gamma(\phi\rightarrow K\overline{K})$
 & $\left[M_{\rho\omega}\right]_{\rm nonEM}$ &
 & $[\Gamma(\omega)/\Gamma(\rho)]_{\pi\pi}$
 & $\left[\Delta m_{K^*}\right]_{\rm nonEM}$
 & $\left[\mbox{Br}(\phi)\right]_{\pi\pi}$
 \\
 $x$ & (MeV)  & (MeV)  & $g_{\omega\pi\pi}$
 & $\times10^{-3}$ & (MeV)
 & $\times10^{-4}$ \\
\hline\hline
$14$ & $3.52$ & $-2.51$ & $-0.0525$ & $0.95$ & $2.18$ & $0.717$ \\
$17$ & $3.52$ & $-2.63$ & $-0.0550$ & $1.06$ & $2.29$ & $0.78$ \\
$20.5$ & $3.52$ & $-2.81$ & $-0.0585$ & $1.24$ & $2.45$ & $0.869$ \\
$23$ & $3.52$ & $-2.96$ & $-0.0616$ & $1.41$ & $2.59$ & $0.944$ \\
$26$ & $3.52$ & $-3.18$ & $-0.0659$ & $1.65$ & $2.79$ & $1.05$ \\
$29$ & $3.52$ & $-3.44$ & $-0.0709$ & $1.97$ & $3.02$ & $1.17$ \\
$32$ & $3.53$ & $-3.75$ & $-0.0769$ & $2.39$ & $3.31$ & $1.31$ \\
\hline\hline
 expt. & $3.69\pm0.06$ & & & $1.23\pm0.19$
&
&{\footnotesize  $(0.8_{\displaystyle-0.4}^{\displaystyle+0.8})$}
 \\
\hline
\end{tabular}
\end{center}
\caption[]{Predictions.}
\label{tab: predictions 1}
\end{table}

First consider the prediction for
$\Gamma(\phi\rightarrow K\overline{K})$.
It is seen to be in reasonably good agreement with experiment
and also to be essentially independent of $x$.
This can be understood by noting that
in the case $\nu_a=\nu_b=0$,
we have the following
relation among $\rho\phi\phi$ couplings:
\ba
g_{\phi {\scriptscriptstyle K\overline{K}}} &=&
2
g_{{\scriptscriptstyle K^{\ast}K}\pi}
\left(\frac{F_{\pi p}}{F_{Kp}}\right)
\left(\frac{Z_{K^{\ast}}}{Z_\phi}\right)
- g_{\rho\pi\pi}
\left(\frac{F_{\pi p}}{F_{Kp}}\right)^2
\left(\frac{Z_\rho}{Z_\phi}\right)
\ .
\label{rel: couplings}
\ea
Each ratio of two $Z$'s is independent of parameter
choice, as discussed below eq.~(\ref{mass rel vector 2}).
Then we obtain $\Gamma(\phi\rightarrow\pi\pi)=3.71$(MeV)
independently of $x$ for $\nu_a=\nu_b=0$.
When we include the small OZI rule violating terms,
this relation is only slightly changed.

Next consider the $\omega\rightarrow\pi\pi$ process.
Details of the calculation are given in
Appendix~\ref{app: omega pi pi}.
Examining table~\ref{tab: predictions 1} shows that the result depends
sensitively on $x$ and that
the experimental value
$\Gamma(\omega\rightarrow\pi\pi)/
\Gamma(\rho\rightarrow\pi\pi)=1.23$ selects
\be
x = 20.5 \ , \mq
R = 48.2 \ , \mq
\left[M_{\rho\omega}\right]_{\rm nonEM} = - 2.81 \, \mbox{(MeV)}
\ .
\label{Rval 1}
\ee
Now from table~\ref{tab: parameters} we read off $y=-0.202$ and hence
that the fundamental quark masses are estimated to stand in the ratio
\be
m_u \, : \, m_d \, : \, m_s \,\, = \,\,
1 \, : \, 1.51 \, : \, 25.7  \ .
\label{quark mass ratio}
\ee
It turns out that the effect of the direct $\omega\rightarrow2\pi$
vertex, which had not been considered in previous discussions, is
small.
This may be understood in the following way.
If we neglect the small effect from $\gamma'$ in $M_{\rho\omega}$
and the small deviation from the relation
$g_{\rho\pi\pi}=m_\rho^2/(\gtil F_{\pi p}^2)$,
the ratio of the mass mixing and direct contributions is
estimated in terms of the ratio of $\rho$
meson mass to width as
\be
\left(\frac{M_{\rho\omega}}{\Gamma_\rho/2}\right)^2 \, : \,
\left(\frac{g_{\omega\pi\pi}}{g_{\rho\pi\pi}}\right)^2
\simeq
\left(\frac{-4y\alpha_+}{m_\rho\Gamma_\rho}\right)^2 \, : \,
\left(\frac{-4y\alpha_+}{m_\rho^2}\right)^2
\simeq
\left( \frac{m_\rho}{\Gamma_\rho} \right)^2 \, : \, 1
\simeq 26 : 1 \ .
\ee
[Using the values shown in table~\ref{tab: predictions 1}
we find
$(2M_{\rho\omega}/\Gamma_\rho)^2\, : \,
(g_{\omega\pi\pi}/g_{\rho\pi\pi})^2
\simeq 30 \, : \, 1$.]
In the present model
the following relation between vector meson masses
and $\rho$--$\omega$ mixing is satisfied
when the OZI violating $\nu$ terms are neglected:
\be
\left[M_{\rho\omega}\right]_{\rm nonEM}
= - \frac{1}{R} \frac{m_{K^\ast}^2-m_\rho^2}{2m_\rho} \,
\left(\frac{Z_{K^{\ast}}}{Z_\rho}\right)^2
\ .
\label{rel: rho omega}
\ee
This is similar to the relation derived in
ref.~\cite{Gasser-Leutwyler:PRep}.
Thus our value in eq.~(\ref{Rval 1}) is close to the value in
ref.~\cite{Gasser-Leutwyler:PRep}.

{}From table~\ref{tab: predictions 1} we observe that the
predicted branching ratio
$\Gamma(\phi\rightarrow\pi\pi)/
\Gamma(\phi\rightarrow\mbox{\rm all})$
(see Appendix~\ref{app: omega pi pi} for details)
for $x=20.5$ agrees well with the central experimental value.
However the experimental error is large.

Our final prediction of this type is seen from
table~\ref{tab: predictions 1} to be
\be
\left[\Delta m_{K^{\ast}} \right]_{\rm nonEM} = 2.45
\ \mbox{\rm MeV} \ .
\ee
This number may be roughly
understood from the predicted ratio which
holds with $\nu_a=\nu_b=0$:
\be
\frac{\left[\Delta m_{K^{\ast}}\right]_{\rm nonEM}}%
{-\left[M_{\rho\omega}\right]_{\rm nonEM}}
=\frac{m_\rho}{m_{K^{\ast}}}
\left( \frac{Z_\rho}{Z_{K^{\ast}}}\right)^4
\simeq 1.0
\ .
\label{rel: rho omega del K}
\ee
On the other hand,
the Particle Data Group\cite{ParticleDataGroup:94} tells us that
\be
\Delta m_{K^{\ast}} =
\left\{
\begin{array}{l@{\qquad\qquad}l}
4.5\pm0.4 \ \mbox{\rm MeV} & \mbox{\rm (a)} \\
6.7\pm1.2 \ \mbox{\rm MeV} & \mbox{\rm (b)} \ ,
\end{array}
\right.
\label{eq: value K}
\ee
depending on whether one (a) simply subtracts the listed $K^{\ast+}$
mass from the listed $K^{\ast0}$ mass or
(b) considers just the ``dedicated'' experiments.
In order to compare this with our prediction it is necessary to take
the electromagnetic piece
$\left[\Delta m_{K^{\ast}}\right]_{\rm EM}$,
defined by
\be
\Delta m_{K^{\ast}} =
\left[\Delta m_{K^{\ast}}\right]_{\rm nonEM} +
\left[\Delta m_{K^{\ast}}\right]_{\rm EM}
\ee
into account.
A bag model estimate\cite{Deshpande-Dicus-Johnson-Teplitz}
gave $\left[\Delta m_{K^{\ast}}\right]_{\rm EM}=-0.7$MeV.
Thus, and this is an old puzzle,
there appears to be a serious disagreement.
In a previous paper\cite{Schechter-Subbaraman-Weigel},
this motivated a compromise fit in which $x$ was much larger
(since $\left[\Delta m_{K^{\ast}}\right]_{\rm nonEM}$ increases
with increasing $x$) but $m(\phi)$ was allowed to differ somewhat from
its experimental value.
In the next subsection we will investigate the effect of the
${\cal M}^2$,
$\mu$--type terms on this puzzle.

Now, however, we would like to see what a ``model-independent''
estimate for $\left[\Delta m_{K^{\ast}}\right]_{\rm EM}$
has to say on this matter.
For the pseudoscalars one usually employs such an
approach\footnote{
In our present language,
Dashen's theorem amounts to the observation that the only non
derivative effective operator which agrees with the chiral
transformation property of the product of two electromagnetic
currents is $\widetilde{U}_{11}\widetilde{U}_{11}^{\dag}
= (4/F_\pi^2) (\pi^+\pi^- + K^+K^-) + \cdots$.
This gives $\Delta m_\pi^\gamma = \Delta m_K^\gamma$.
Here $U = U_0 \widetilde{U}$ with
$\mbox{\rm det} \widetilde{U} = 1$.
}
(i.e., Dashen's theorem\cite{Dashen})
to estimate the analogous quantity
$\left[\Delta m_K\right]_{\rm EM}$.
For the vectors we may derive a sum rule for the electromagnetic parts
of the mass differences by assuming the OZI rule in the form that the
vector field appears as a nonet and the effective operator is a single
trace.
Corresponding to photon exchange we should have two powers of $Q$
(defined in eq.~(\ref{def: chrages})).
The effective operator describing the electromagnetic contributions to
the mass splittings then takes the form
\be
A \, \mbox{Tr}\, \left( Q \rho_\mu Q \rho_\mu \right)
+ B \, \mbox{Tr}\, \left( Q^2 \rho_\mu \rho_\mu \right)
\ ,
\ee
where $A$ and $B$ are some constants.
This leads to the sum rule:
\be
\left[\Delta m_{K^{\ast}}\right]_{\rm EM} =
\left[ m(K^{\ast0}) - m(K^{\ast+}) \right]_{\rm EM}
=
\frac{m_\rho}{m_{K^{\ast}}}
\Biggl[
 \left[m(\rho^0) - m(\rho^+) \right]_{\rm EM}
 - \left[M_{\rho\omega}\right]_{\rm EM}
\Biggr] \ .
\ee
$\left[M_{\rho\omega}\right]_{\rm EM}$ has
(see eq.~(\ref{eq: rho - omega from photon}))
been estimated as $0.42$MeV.
Furthermore, since $m(\rho^0)-m(\rho^+)$ is a $\Delta I = 2$
object,
it is a good approximation to neglect the quark mass
($\Delta I=1$ at first order) contribution and set
\be
\left[ m(\rho^0) - m(\rho^+) \right]_{\rm EM}
\simeq
\left[ m(\rho^0) - m(\rho^+) \right]
= 0.3 \pm 2.2 \, \mbox{\rm MeV} \ ,
\ee
where the PDG\cite{ParticleDataGroup:94} estimate was used in the last
step.
We thus get the desired estimate
\be
\left[\Delta m_{K^{\ast}}\right]_{\rm EM} =
-0.11 \pm 1.9 \, \mbox{\rm MeV} \ .
\ee
With our prediction for
$\left[\Delta m_{K^{\ast}}\right]_{\rm nonEM}$
we would then have
\be
\Delta m_{K^{\ast}} = 2.34 \pm 1.9 \, \mbox{\rm MeV} \ .
\ee
This value has a large enough uncertainty to possibly agree with the
experimental value in eq.~(\ref{eq: value K}.a) above.
In any event,
the large uncertainty associated with the $\rho^+$--$\rho^0$
experimental mass splitting suggests a certain skepticism
about accepting literally the stated $K^{\ast0}$--$K^{\ast+}$
experimental value.

\subsection{Second Order Effects
\label{sec: second order}}

\begin{figure}[htbp]
\begin{center}
\ \epsfbox{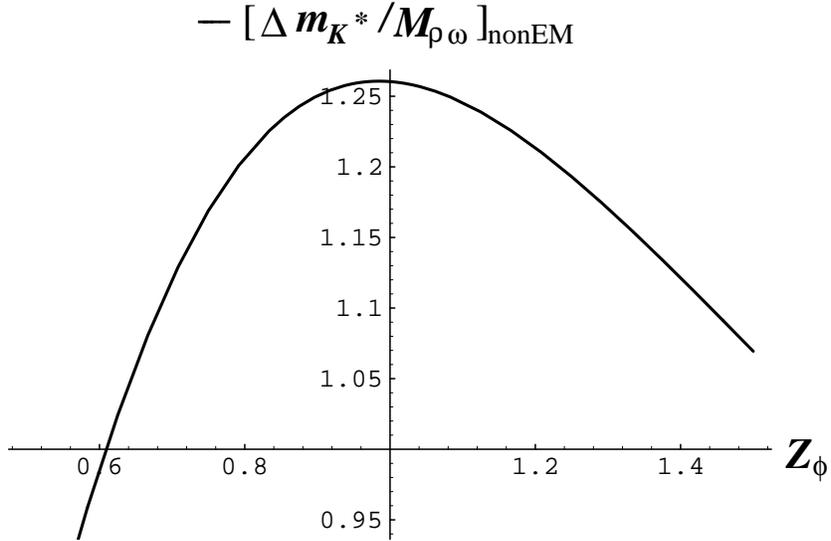}
\end{center}
\caption[]{The dependence of the ratio
$\left[\Delta m_{K^{\ast}}/M_{\rho\omega}\right]_{\rm nonEM}$
on $Z_\phi$ for $x=20.5$.
We use $\theta_{\phi\omega}=0.055$.
}
\label{fig: ratio 1}
\end{figure}

In this section we study the effects of the $\mu$-labeled,
${\cal M}^2$ symmetry breakers for the vector meson nonet.
These are given in eq.~(\ref{Lag: second}) and
rewritten in terms of convenient linear combinations in
eq.~(\ref{eq: mu terms 2}).
(Note that the formulas listed in Appendix~\ref{app: formula}
contain the $\mu$-term contributions.)
The motivation for this study is to see if we can fit the model to a
larger value for $\left[\Delta m_{K^{\ast}}\right]_{\rm nonEM}$,
which could be required by future precision experiments.
Without the $\mu$-terms (and neglecting the OZI violating
$\nu$-terms), eq.~(\ref{rel: rho omega del K}) shows that
$\left[\Delta m_{K^{\ast}}\right]_{\rm nonEM}$
can not be any greater than
$-\left[M_{\rho\omega}\right]_{\rm nonEM}$ when all the input
parameters are fit exactly.
We now have a similar relation in the presence of the $\mu$-terms.
Still keeping $\nu_a=\nu_b=0$
and
using the formulae given in Appendix~\ref{app: formula}
we find
\be
\left[\frac{\Delta m_{K^{\ast}}}{-M_{\rho\omega}}\right]_{\rm nonEM}
= \frac{m_\rho}{m_{K^{\ast}}} \times
\frac{m_\phi^2-m_{K^{\ast}}^2}{m_{K^{\ast}}^2-m_\rho^2} \times
\frac{Z_\rho^2 Z_\phi^2}{Z_{K^{\ast}}^4} < 1.0 \ .
\label{eq: second ratio 1}
\ee
To get the final inequality,
we used
$(Z_\rho Z_\phi)/(Z_{K^{\ast}}^2) =
2(Z_\rho Z_\phi)/(Z_\rho^2 + Z_\phi^2) \leq 1$.
When we include the OZI violating $\nu_a$ and $\nu_b$ terms,
the relation (\ref{eq: second ratio 1})
is slightly changed to:
\be
\left[
\frac{\Delta m_{K^{\ast}}}{-M_{\rho\omega}}
\right]_{\rm nonEM}
= \frac{m_\rho}{m_{K^{\ast}}} \times
\frac{Z_\rho^2}{Z_{K^{\ast}}^2} \times
\frac{Z_\phi^2(m_\phi^2-m_{K^{\ast}}^2)
+ Z_\rho^2(m_\omega^2-m_\rho^2)/2
- Z_\rho Z_\phi (\sqrt{2}\Pi_{\phi\omega})}
{Z_{K^{\ast}}^2(m_{K^{\ast}}^2-m_\rho^2)
- Z_\rho^2(m_\omega^2-m_\rho^2)/2
+ Z_\rho Z_\phi (\sqrt{2}\Pi_{\phi\omega})/2} \ .
\label{eq: second ratio 2}
\ee
In order to evaluate this expression we should fit the Lagrangian
parameters including the $\mu$ terms.
We shall set $\mu_c=0$ since it only affects the pseudoscalar decay
constants.
$\mu_b$ and $\mu_d$ do not contribute to the quantities discussed  in
this paper.
We shall fit $\mu_a$ and $\mu_e$ to the experimental
values of the
$\phi\rightarrow K\overline{K}$ and $\omega\rightarrow\pi\pi$ decay
widths
(these two quantities are thus no longer predictions).
The quark mass ratio $x$ will be kept fixed at $20.5$.
It is then convenient to consider parameters
evaluated for various
values of $Z_\phi$ (see eq.~(\ref{def: Zrho})).
These are shown in table~\ref{tab: second order}.
\begin{table}[htbp]
\begin{center}
\small
\begin{tabular}{|c||c|c|c|c|c|c||c|c|c|}
\hline
 & $\mu_a$ & $\mu_e$ & $m_v$ & & $\alpha_+$ & $\alpha_-$
 & \multicolumn{3}{c|}{$V\rightarrow e^+e^-$}
\\
\cline{8-10}
 $Z_\phi$ & $\times10^{-3}$ & $\times10^{-3}$ & (GeV)
 & $\widetilde{g}$ & $\times10^{-3}$ & $\times10^{-3}$
 & $\Gamma(\rho)$ & $\Gamma(\omega)$ & $\Gamma(\phi)$
\\
\hline\hline
$0.5$ & $0.204$ & $-0.00227$ & $0.761$ & $4.15$
 & $2.01$ & $2.75$ & $4.80$ & $0.583$ & $0.314$ \\
$0.8$ & $0.0994$ & $-0.0186$ & $0.759$ & $4.10$
 & $-1.78$ & $1.08$ & $5.03$ & $0.638$ & $0.875$ \\
$1.0$ & $-0.000542$ & $0.00417$ & $0.756$ & $4.03$
 & $-5.28$ & $-0.676$ & $5.28$ & $0.689$ & $1.44$ \\
$1.1$ & $-0.0595$ & $0.0241$ & $0.755$ & $3.99$
 & $-7.33$ & $-1.80$ & $5.44$ & $0.720$ & $1.79$ \\
$1.5$ & $-0.355$ & $0.153$ & $0.747$ & $3.75$
 & $-17.4$ & $-8.25$ & $6.49$ & $0.905$ & $3.85$ \\
\hline
\end{tabular}
\end{center}
\caption[]{The dependences of the parameters on $Z_\phi$
for $x=20.5$.
The units of the parameters
$\mu_a$, $\mu_e$, $\alpha_+$ and $\alpha_-$
are (GeV)$^2$.
The predictions for $V\rightarrow e^+e^-$ are also shown.
[The units are (KeV).]
}
\label{tab: second order}
\end{table}
We next show the dependence of
$\left[\Delta m_{K^{\ast}}/M_{\rho\omega}\right]_{\rm nonEM}$
on $Z_\phi$ for $x=20.5$ in fig.~\ref{fig: ratio 1}.
Here we use as an input, $\theta_{\phi\omega}$
as derived from $\phi\rightarrow\pi^0\gamma$ decay width
for $x$ around $20.5$ in section~\ref{sec: OZI violation}.
It is clear from Fig.~\ref{fig: ratio 1} that
$\left[\Delta m_{K^{\ast}}\right]_{\rm nonEM}$ is still approximately
bounded by $-\left[M_{\rho\omega}\right]_{\rm nonEM}$.
If a more precise experimental value of
$\Delta m_{K^{\ast}}$ turns out in the $5$ KeV range it would be
difficult to explain in this model as arising from
$\left[\Delta m_{K^{\ast}}\right]_{\rm nonEM}$ --
a suitably large value of
$\left[\Delta m_{K^{\ast}}\right]_{\rm EM}$ would be required.

In table~\ref{tab: second order}
we also show predictions for the $\rho^0$, $\omega$ and $\phi$ decay
widths into $e^+e^-$.
These will be discussed in more detail in the next section.
Here we just want to point out that getting agreement with experiment
does in fact require $Z_\phi$ to be within about $\pm25$\%
of unity.

\newpage

\msection{Electroweak Processes}

In the previous section we determined the parameters of the effective
Lagrangian for various values of the quark mass ratio, $x$.
Some predictions were also made and a best value of $x$ around $20$
was selected.
All the physical quantities discussed were independent of the
electroweak gauge fields.
In this section we will study processes which are related to the
electroweak interaction.
The electroweak gauge fields may be introduced, without introducing
any new arbitrary constants, according to the prescription of
(\ref{eq: replacement})--(\ref{def: chrages}).
Using our Lagrangian, we will first give predictions for
the $V\rightarrow l^+l^-$ ($V=\rho$, $\omega$, $\phi$;
$l=e$, $\mu$) decays.
Next, we calculate form factors;
namely
the pion and kaon charge radii as well as the
the slope parameter of $K_{e3}$ decay.
Finally, we will discuss a direct fit of the experimental
$e^+e^- \rightarrow \pi^+\pi^-$ data using our Lagrangian.
This provides a consistency check as well as an indication of what is
involved in obtaining vector meson parameters from experiment.

\subsection{$V\rightarrow e^+e^-$ decay processes
\label{sec: V -> ee}}

Let us write the effective Lagrangian describing
the $V$--$\gamma$ transition terms in our model as:
\be
\Lag_{V\gamma} \equiv
e {\cal A}_\mu
\left[
g_\rho \rho_\mu + g_\omega \omega_\mu +g_\phi \phi_\mu
\right]
\ .
\label{eq: V - gamma mixing}
\ee
The expressions for the transition strengths
$g_\rho$, $g_\omega$ and $g_\phi$
are given in eq.~(\ref{eq: mixing strengths}).
In terms of these transition strengths
the $V\rightarrow l^+ l^-$ decay widths are given by
\be
\Gamma(V\rightarrow l^+l^-) =
\frac{4 \pi \alpha^2}{3 }
\abs{\frac{g_{Vp}}{M_V^2}}^2
\frac{M_V^2+2m_l^2}{M_V^2} \sqrt{M_V^2-4m_l^2} \ , \
\label{V ee width}
\ee
where
\ba
g_{\rho_p} = g_\rho \  , \qquad
g_{\omega_p} =
g_\omega - \theta_{\phi\omega} g_\phi \ , \qquad
g_{\phi_p} =
g_\phi + \theta_{\phi\omega} g_\omega \ .
\label{def: Gtil}
\ea
Here we included the $\phi$--$\omega$ mixing from
the
OZI rule violating terms $\Lag_{\nu1}+\Lag_{\nu2}$.

Before calculating the partial widths,
we note some relations between the ratios
of these partial widths
and the vector meson masses.
In the case where $\nu_a=\nu_b=0$,
the relation
\be
\frac{g_\rho}{m_\rho^2} = 3 \frac{g_\omega}{m_\omega^2}
= -\frac{3}{\sqrt{2}}\frac{Z_\rho}{Z_\phi}
\frac{g_\phi}{m_\phi^2}
= \frac{Z_\rho}{\sqrt{2}\gtil}
\ee
is satisfied.
Noting that the values of $Z_\rho$ and $Z_\phi$ depend on $x$
while their ratio,
according to the discussion around eq.~(\ref{mass rel vector 2}),
depends only on masses
we find the following $x$-independent relations:
\be
\frac{g_\omega}{g_\rho} = \frac{1}{3}
\frac{m_\omega^2}{m_\rho^2}\ ,
\qquad
- \frac{g_\phi}{g_\omega} =
\sqrt{2}\frac{m_\phi^2}{m_\rho^2} \frac{Z_\phi}{Z_\rho}
\simeq 2.2 \ .
\ee
{}From these relations
we obtain
$\Gamma(\omega\rightarrow e^+e^-)/\Gamma(\rho\rightarrow e^+e^-)
\simeq0.11$ and
$\Gamma(\phi\rightarrow e^+e^-)/\Gamma(\omega\rightarrow e^+e^-)
\simeq2.2$.
These ratios are in reasonable agreement with the experimental values
given by $0.09$ and $2.3$, respectively.

Now, including the OZI violation terms
we calculate the decay widths
using the parameters determined in the previous section
and display them in
table~\ref{tab: V -> ee decay}.
\begin{table}[htbp]
\begin{center}
\small
\begin{tabular}{|c||c|c|c||c|}
\hline
 & \multicolumn{4}{|c|}{$V\rightarrow e^+e^-$} \\
\hline
 $x$ & $\Gamma(\rho)$ & $\Gamma(\omega)$  & $\Gamma(\phi)$ &
 $\Gamma(\rho)_{\rm GS}$$^{\dag}$
 \\
\hline\hline
$14$   & $5.28$ & $0.688$ & $1.37$ & $6.31$ \\
$17$   & $5.26$ & $0.685$ & $1.37$ & $6.29$ \\
$20.5$ & $5.24$ & $0.682$ & $1.36$ & $6.27$ \\
$23$   & $5.23$ & $0.680$ & $1.36$ & $6.26$ \\
$26$   & $5.23$ & $0.678$ & $1.35$ & $6.25$ \\
$29$   & $5.22$ & $0.676$ & $1.35$ & $6.24$ \\
$32$   & $5.21$ & $0.673$ & $1.35$ & $6.24$ \\
\hline\hline
 exp. &
 $6.77\pm0.32$ & $0.60\pm0.02$ & $1.37\pm0.05$
 & $6.77\pm0.32$ \\
\hline
\end{tabular}
\end{center}
\caption[]{
The predictions for $V\rightarrow l^+l^-$ decay widths
in units of KeV.
$\Gamma(\rho)_{\rm GS}$ indicates that
the Gounaris-Sakurai\cite{Gounaris-Sakurai} effect is included
(see text).
}
\label{tab: V -> ee decay}
\end{table}
It is seen that the $x$-dependences are quite small.
The predicted value for the $\phi$ decay agrees well
with experiment,
while
the $\omega$ meson prediction is about 10\% too high and the $\rho$
meson prediction is about 25\% too low.
Let us consider the last point in detail.
In our model, the ratio of $\Gamma(\rho\rightarrow e^+e^-)$
to $\Gamma(\rho\rightarrow \pi^+\pi^-)$ is given by
\be
R_{e\pi} \equiv
\frac{\Gamma(\rho\rightarrow e^+e^-)}%
{\Gamma(\rho\rightarrow \pi^+\pi^-)}
= \frac{\alpha^2}{36}
\left\vert\frac{m_\rho^2 - 4m_\pi^2}{m_\rho^2}\right\vert^{3/2}
\left(\frac{m_\rho}{\Gamma_\rho}\right)^2
\left( \frac{g_\rho g_{\rho\pi\pi}}{\sqrt{2}m_\rho^2} \right)^2
\ .
\label{eq: ee pipi ratio}
\ee
Since the contribution from the $\alpha_-$ term to
$g_{\rho\pi\pi}$ is very small,
the KSRF relation $2 m_\rho^2 = g_{\rho\pi\pi}^2 F_\pi^2$\footnote{
Experimentally,
$(2m_\rho^2)/(g_{\rho\pi\pi}^2F_{\pi p}^2) \simeq 0.957$.
This form of the KSRF relation agrees with the
choice $k=2$ in
eq.~(\ref{def: k}) when symmetry breaking is neglected.
}
means
the last factor in eq.~(\ref{eq: ee pipi ratio})
is very close to unity,
i.e.,
$g_\rho\simeq\sqrt{2}m_\rho^2/g_{\rho\pi\pi}$.
Then the predicted value of the
$\rho\rightarrow e^+e^-$ decay width
is seen to be somewhat
small compared with the experimental value.

A natural way to fine tune our predictions is to include gauge
invariant, higher derivative
$\rho$--$\gamma$ transition terms.
Let us introduce the following effective Lagrangian terms
which describe the ``kinetic type'' $\rho$--$\gamma$
transitions:
\ba
\Lag_{V\gamma} &=&
h  \kappa_0 \Tr \Biggl[
F_{\mu\nu}(\rho)
\left\{
\xi F^R_{\mu\nu}(B) \xi^{\dag}
+ \xi^{\dag} F^L_{\mu\nu}(B) \xi
\right\}
\Biggr]
\nonum
&& \,
+ h \frac{\kappa_1}{2}
\Tr \Biggl[
\left\{ \mhatp , F_{\mu\nu}(\rho) \right\}
\left(
\xi F^R_{\mu\nu}(B) \xi^{\dag}
+ \xi^{\dag} F^L_{\mu\nu}(B) \xi
\right)
\Biggr]
\nonum
&& \,
+ h \frac{\kappa_2}{2}
\Tr \Biggl[
\left[ \mhatm , F_{\mu\nu}(\rho) \right]
\left(
\xi F^R_{\mu\nu}(B) \xi^{\dag}
- \xi^{\dag} F^L_{\mu\nu}(B) \xi
\right)
\Biggr]
\nonum
&& \,
+ h \kappa_3
\Tr \left[ F_{\mu\nu}(\rho) \right]
\Tr \left[ F^R_{\mu\nu}(B) + F^L_{\mu\nu}(B) \right]
\nonum
&& \,
+ h \kappa_4
\Tr \left[ \mhatp F_{\mu\nu}(\rho) \right]
\Tr \left[ F^R_{\mu\nu}(B) + F^L_{\mu\nu}(B) \right]
\nonum
&& \,
+ h \kappa_5
\Tr \left[ F_{\mu\nu}(\rho) \right]
\Tr \Biggl[
\mhatp
\left(
\xi F^R_{\mu\nu}(B) \xi^{\dag}
+ \xi^{\dag} F^L_{\mu\nu}(B) \xi
\right)
\Biggr]
\label{eq: lag V-g }
\ea
where $\widehat{\cal M}_{\pm}$
is defined
in eq.~(\ref{def: mass p m}).
The $\kappa_0$ term is the leading kinetic type mixing term
while the $\kappa_1$ and $\kappa_2$ terms
describe the first order symmetry breaking
pieces which conserve the OZI rule.
The $\kappa_3$ term describes the OZI rule violating contribution to
the $\rho$--$\gamma$ mixing and the first order symmetry breaking
and OZI rule violating pieces are given by the
$\kappa_4$ and $\kappa_5$ terms.
Noting that
$\Tr(Q)=0$
and that
$\mhatm$ must include at least one pseudoscalar meson,
we see that only
the $\kappa_0$, $\kappa_1$ and $\kappa_5$ terms contribute to the
kinetic type $\rho$--$\gamma$ mixing
without pseudoscalar mesons.
Let the $C_V$
denote kinetic type $\rho$--$\gamma$ mixing coefficients
in the effective Lagrangian:\footnote{
In ref.~\cite{OConnel-etal:rho-omega}
such a term has been advocated to fit all the $V$--$\gamma$ mixing,
rather than just the corrections to eq.~(\ref{eq: V - gamma mixing}).
}
\be
\Lag_{V\gamma\,{\rm kin}}
= e (\delm {\cal A}_\nu - \del_\nu {\cal A}_\mu ) \left[
 C_\rho \delm \rho^0_\nu + C_\omega \delm \omega_\nu
 + C_\phi \delm \phi_\nu
\right] \ .
\ee
The $C_V$ are then given by
\ba
C_\rho &=&
\frac{4}{\sqrt{2}Z_\rho}\left[
\kappa_0 + \kappa_1 \right] \ ,
\nonum
C_\omega &=&
\frac{4}{3\sqrt{2}Z_\omega}\left[
\kappa_0 + \kappa_1 - 2(x-1) \kappa_5 \right] \ ,
\nonum
C_\phi &=&
- \frac{4}{3Z_\phi}\left[
\kappa_0 + x \kappa_1 + (x-1) \kappa_5 \right] \ .
\label{def: kinetic mixing}
\ea
These kinetic mixing terms are included by replacing
$g_{\scriptscriptstyle V_p}$ by
$g_{\scriptscriptstyle V_p}- C_{V_p}M_V^2$
in eq.~(\ref{V ee width}),
where we also take the $\phi$--$\omega$ mixing into account.

We determine the values of these parameters from the experimental
$e^+e^-$ decay widths and obtain
\be
C_\rho \simeq - 0.024 \ , \qquad
C_\omega \simeq 0.0039 \ , \qquad
C_\phi \simeq 0.00005 \ ,
\label{fit: kinetic mixing}
\ee
for $x=20.5$.
[The variations of $C_\rho$ and $C_\omega$ are less than 5\%
for a wide range of $x$-values ($14\leq x\leq45$).
The range of $C_\phi$ is $-0.00027\sim0.001$ for $14\leq x\leq45$.]
The large difference between $C_\rho$ and $C_\omega$ implies that
the $\kappa_5$ term gives a contribution
which is comparable to the leading order one; $2x\kappa_5\sim
\kappa_0$.
The $\kappa_5$ term is expected to be suppressed by the OZI rule in
addition to the smallness
of the SU(3) symmetry breaking mass.
A similar effect was observed in section~\ref{sec: OZI violation}.
We should notice that the main contribution to the $\rho$--$\gamma$
transition is supplied by the mass type mixing
$g_{\scriptscriptstyle V}$,
the value of which is determined from the pure hadronic sector and
is almost consistent with the experiment.
In other words,
the plausible KSRF relation
$g_\rho=\sqrt{2}m_\rho^2/g_{\rho\pi\pi}$ is naturally explained by the
mass type mixing, and small corrections are given by the kinetic type
mixing.
For the $\rho$ meson case,
the kinetic type mixing gives about a 15\% correction to the
mass type mixing amplitude in order to achieve agreement with
experiments.

Another possibility for improving the
$\Gamma(\rho^0\rightarrow e^+e^-)$ prediction
is
to take the large $\rho$ width into account,
as Gounaris and Sakurai
pointed out long ago\cite{Gounaris-Sakurai}.
They included the finite width corrections based on a generalized
effective range formula for pion-pion scattering and
got a non-negligible enhancement factor for the pion form factor:
\be
F_\pi(s) =
\frac{m_\rho^2 + d m_\rho \Gamma_\rho}%
{(m_\rho^2-s) + \Pi(s) - i m_\rho \Gamma_\rho(k/k_\rho)^3(m/\sqrt{s})}
\ ,
\label{G-S form}
\ee
where
\ba
&&
\Pi(s) = \Gamma_\rho \frac{m_\rho^2}{k_\rho^3}
\left[
  k^2 \left\{ h(s) - h(m_\rho^2) \right\}
  + k_\rho^2 h'(m_\rho^2) (m_\rho^2-s)
\right] \ ,
\nonumber\\
&&
h(s) = \frac{2}{\pi} \frac{k}{\sqrt{s}}
\ln \left( \frac{\sqrt{s} + 2 k}{2m_\pi} \right) \ ,
\qquad
k = \sqrt{\frac{s - 4 m_\pi^2}{4}} \ ,
\qquad
k_\rho = \sqrt{\frac{m_\rho^2 - 4 m_\pi^2}{4}} \ ,
\nonumber\\
&&
d =
\frac{3}{\pi}\frac{m_\pi^2}{k_\rho^2}
\ln \left( \frac{m_\rho + 2 k_\rho}{2m_\pi} \right)
+ \frac{m_\rho}{2\pi k_\rho}
- \frac{m_\pi^2m_\rho}{\pi k_\rho^3} \ .
\label{G-S parts}
\ea
As a result the above ratio (\ref{eq: ee pipi ratio})
is enhanced to
\be
\left. R_{e\pi} \right\vert_{\rm finite\  width}
=
R_{e\pi} \times
\left[ 1 + d \left( \Gamma_\rho/ m_\rho \right) \right]^2 \ .
\ee
In our model this effect is included by the replacement
\be
g_{\rho} \rightarrow g_\rho \times
\left[ 1 + d \left( \Gamma_\rho/ m_\rho \right) \right] \ ,
\label{g rho : GS effect}
\ee
which enhances the predicted value of
$\Gamma(\rho\rightarrow e^+e^-)$ by about 20\%.
We show the result of this enhancement in the fifth column in
table~\ref{tab: V -> ee decay}.
Clearly, this G-S effect improves the prediction.
It should be noticed that
the inclusion of the pion loop correction
to the $\rho$ propagator and to the
$\rho\pi\pi$ coupling with a suitable
(vector) on-shell-like renormalization
(we need a higher derivative counter term)
gives the same result as the G-S form factor.

\subsection{Charge radii and the slope factor of $K_{e3}$
\label{sec: charge radii}}

We now consider the pion and kaon charge radii and
also the slope factor of $K_{e3}$ decay.
In our model
the electromagnetic charge radius
of the pseudoscalar $P$ is expressed as
\be
\left\langle r_P^2 \right\rangle
= \frac{6}{\sqrt{2}}
\sum_V
\frac{g_{\scriptscriptstyle V_p P P}g_{\scriptscriptstyle V_p}}%
{M_V^4} \ ,
\label{eq: charge radii}
\ee
where $V=(\rho,\omega,\phi)$.
When we include the kinetic type $\rho$--$\gamma$ mixing,
the above $g_{\scriptscriptstyle V_p}$'s are replaced with
$g_{\scriptscriptstyle V_p}-C_{V_p}M_V^2$.
The slope factor of $K_{e3}$ decay
is defined by the
linear energy dependence of the
form factor $f_+$ in
the matrix element of $K\rightarrow\pi l\nu$ decay:
\ba
&&
{\cal M} \propto f_+(t)
\left[
  (p_K+p_\pi)_\mu \bar{l} \gamma_\mu (1+\gamma_5)\nu
\right]
+ f_-(t) m_l \bar{l}(1+\gamma_5)\nu \ ,
\nonumber\\
&&
f_+(t) = f_+(0) \left[ 1 + \lambda_+ (t/m_\pi^2)\right] \ .
\ea
In our model this form factor is dominated by the $K^*$ meson exchange
diagram.
Then $\lambda_+$ is expressed as
\be
\frac{\lambda_+}{m_\pi^2} =
\frac{2F_{Kp} F_{\pi p}}{F_{Kp}^2 + F_{\pi p}^2}
\frac{g_{\scriptscriptstyle K^*}g_{{\scriptscriptstyle K^*K}\pi}}%
{\sqrt{2}m_{K^*}^4} \ ,
\label{def: Ke3 slope}
\ee
where $g_{\scriptscriptstyle K^*} =
\sqrt{2}Z_{K^{\ast}}m_{K^*}^2/\gtil$
is the $K^{\ast}$--$W$ transition strength
defined by
\be
\Lag_{K^{\ast}W} = \frac{g_2}{4} g_{\scriptscriptstyle K^*}
\left( V_{us} K^{*-}_\mu {\cal W}_\mu^+
+ V_{us}^* K^{*+}_\mu {\cal W}_\mu^-
\right)
\ .
\ee
When we include the kinetic type $\rho$--$\gamma$ transition terms
as given in eq.~(\ref{eq: lag V-g }),
this also gives a kinetic type $K^{\ast}$--$W$ transition term:
\be
\Lag_{K^{\ast}W{\rm kin}} = \frac{g_2}{4} C_{K^*}
\left[
 V_{us}
 \left(
  \partial_\mu {\cal W}_\nu^+ - \partial_\nu {\cal W}_\mu^+
 \right)
 \partial_\mu K^{*-}_\nu
+ V_{us}^*
 \left(
  \partial_\mu {\cal W}_\nu^- - \partial_\nu {\cal W}_\mu^-
 \right)
 \partial_\mu K^{*+}_\nu
\right]
\ ,
\ee
where $C_{K^{\ast}}$ is given by
\be
C_{K^{\ast}} =
\frac{2\sqrt{2}}{Z_{K^{\ast}}}
\left[ 2 \kappa_0 + (x+1) \kappa_1 \right] \ .
\ee
This effect is included in the slope parameter $\lambda_+$
by the replacement
$g_{\scriptscriptstyle K^{\ast}}
\rightarrow g_{\scriptscriptstyle K^{\ast}} - C_{K^{\ast}}
m_{K^{\ast}}^2$ in eq.~(\ref{def: Ke3 slope}).

We show our predictions for the charge radii and
the slope parameter
in table~\ref{tab: charge radius}
together with the existing experimental values.
The dependence on the parameter $x$ is very small
as is the case for
the partial width $\Gamma(V\rightarrow e^+e^-)$.
Thus we show only the prediction for $x=20.5$.
(Actually, the variation of the predictions
for different values of $x$
is less than 0.5\%.)
In the first line
we use the $\rho$--$\gamma$ transition strengths given in
eqs.~(\ref{def: Gtil}) and (\ref{eq: mixing strengths}), while
in the second line we include the kinetic type $\rho$--$\gamma$ mixing
corrections as discussed in the previous section.
Here
we used the values of the $C_V$'s
determined from the $e^+e^-$ partial
decay widths.
It is reassuring
that the corrections discussed
in section~\ref{sec: V -> ee}
to improve $\Gamma(V \rightarrow e^+e^-)$ also improve the predictions
in table~\ref{tab: charge radius}.
\begin{table}[htbp]
\begin{center}
\begin{tabular}{|l|c|c|c|c|}
\hline
& $\left\langle r_\pi^2 \right\rangle$ (fm)$^2$
& $\left\langle r_{K^+}^2 \right\rangle$ (fm)$^2$
& $\left\langle r_{K^0}^2 \right\rangle$ (fm)$^2$
& $\lambda_+$ \\
\hline\hline
 Prediction 1
 & $0.415$ & $0.260$ & $-0.00141$ & $0.0257$ \\
\hline
 Prediction 2
 & $0.472$ & $0.275$ & $-0.0221$ & $0.0264$ \\
\hline
 Prediction 3
 & $0.424$ & $0.263$ & $-0.00467$ & \\
\hline\hline
\small Molzon (1978)\cite{Molzon:r-K:78}
  & & & $-(0.054\pm0.026)$ & \\
\small Dally (1977)\cite{Dally:r-pi:77}
  & $0.31\pm0.04$ & & & \\
\small Dally (1980)\cite{Dally:r-K:80}
  & & $0.28\pm0.05$ &  & \\
\small Dally (1982)\cite{Dally:r-pi:82}
  & $0.439\pm0.030$ & & & \\
\small Amendolia (1984)\cite{Amendolia:r-pi}
  & $0.432\pm0.016$ & & & \\
\small Barkov (1985)\cite{Barkov:85}
  & $0.422\pm0.013$ & & & \\
\small Amendolia (1986)\cite{Amendolia:r-K}
  & & $0.34\pm0.05$ & & \\
\small Amendolia (1986)\cite{Amendolia:NA7}
  & $0.439\pm0.008$ & & & \\
\small Erkal (1987)\cite{Erkal-Olsson}
  & $0.455\pm0.005$ & $0.29\pm0.04$ & & \\
\hline
\small PDG (1994)\cite{ParticleDataGroup:94}
  & & & & $0.0286\pm0.0022$ \\
\hline
\end{tabular}
\end{center}
\caption[]{
Predictions for the charge radii and the $K_{e3}$ slope parameter
$\lambda_+$ with
the existing experimental values.
The values on the second line (Prediction 2) are given
by the inclusion
of the kinetic type $\rho$--$\gamma$ transition strength
as shown in eq.~(\ref{def: kinetic mixing}).
The values in the third line (Prediction 3) include
the enhancement factor
due to the replacement
(\ref{g rho : GS effect 2}).
\label{tab: charge radius}}
\end{table}

When we take the finite $\rho$ width into account,
the pion form factor
is changed as given in eq.~(\ref{G-S form}).
Since
$\Pi(s)$ in eq.~(\ref{G-S parts}) does not affect the form factor
on the vector meson mass shell
while it does near $s=0$,
the replacement (\ref{g rho : GS effect})
is not valid for the charge radii.
Instead, the following replacement
is obtained in the low energy region:
\ba
g_{\rho} &\rightarrow& g_\rho \times
\frac{ 1 + \widetilde{d} \left( \Gamma_\rho/ m_\rho \right) }%
{ 1 + d \left( \Gamma_\rho/ m_\rho \right) } \ ,
\nonumber\\
&&
\widetilde{d} =
\frac{m_\rho^2 + 2m_\pi^2}{2\pi k_\rho^2}
\ln \left( \frac{m_\rho + 2 k_\rho}{2m_\pi} \right)
+ \frac{m_\rho}{2\pi k_\rho}
- \frac{m_\rho^3}{3\pi k_\rho^3} \ .
\label{g rho : GS effect 2}
\ea
We show
in the third line
of table~\ref{tab: charge radius}
the predictions gotten by this replacement.
The predictions are only
slightly improved by the inclusion of the G-S
effect.

Finally
we make a comment on the $K^0$ charge radius and the $\gamma'$
symmetry breaking term in eq.~(\ref{Lag: first}).
Using the first order ($\nu_a=\nu_b=0$) formula
given in appendix~\ref{app: formula},
we find
\ba
\left\langle r_{K^0}^2 \right\rangle &=&
- \frac{8(x+1)}{\gtil^2F_{Kp}^2}
\left[
  \gamma' + \frac{\alpha_-}{2}
  \left( \frac{1}{m_\rho^2} - \frac{1}{m_\phi^2} \right)
\right]
\nonumber\\
&\simeq&
- \frac{8(x+1)\gamma'}{\gtil^2F_{Kp}^2}
\simeq - 0.01 \ (\mbox{fm})^2 \ ,
\ea
where in the second line we
neglected symmetry breaking terms of quadratic order
such as $\alpha_+\times\alpha_-$.
This shows that the inclusion of
the $\gamma'$ term is important for the
charge radius of $K^0$.
However, since the first order mass relation
(\ref{mass rel vector 2}) suppresses $\gamma'$,
we obtain a smaller value than the experimental one.
When we include either the G-S-like enhancement factor
with the replacement
(\ref{g rho : GS effect}),
or the kinetic type mixing $C_V$,
the negative $\rho$ contribution is enhanced and the prediction is
improved
(more substantially in the kinetic mixing case).

\subsection{$e^+e^-\rightarrow\pi\pi$ process
\label{sec: ee -> pi pi}}

In section~\ref{sec: V -> ee}
we determined the value of the coefficient of the kinetic type
$\rho$--$\gamma$ mixing $C_\rho$ from the experimental value of
$\Gamma(\rho\rightarrow e^+e^-)$.
We used the decay width $\Gamma(\omega\rightarrow\pi\pi)$
to fit $x$ in section~\ref{sec: prediction}.
These two decay processes are related to the
$e^+e^-\rightarrow\pi\pi$ process.
In this section we directly fit $C_\rho$ and $x$ from
the experimental data describing the pion form factor of
$e^+e^-\rightarrow\pi\pi$\cite{Barkov:85}
in the energy region $0.73\le\sqrt{s}\le0.83$(GeV).
The restriction of the energy region reduces the dependence on effects
which are difficult to calculate reliably.

\begin{figure}[htbp]
\begin{center}
\ \epsfbox{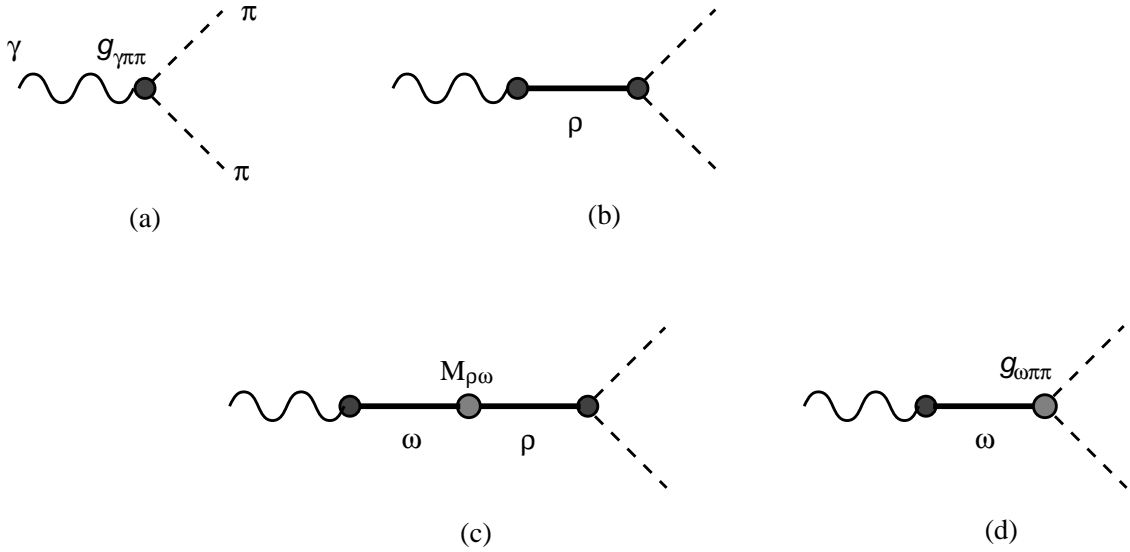}
\end{center}
\caption[]{The Feynman diagrams
contributing to the pion form factor:
(a) the direct $\gamma\pi\pi$ vertex;
(b) the $\rho$ meson exchange diagram;
(c) the $\rho$--$\omega$ mixing;
(d) the direct $\omega\pi\pi$ vertex.
}
\label{fig: ee -> pi pi diagram}
\end{figure}

In the present model
there exist four kinds of contributions to the pion form factor.
We show the corresponding Feynman diagrams in
fig.~\ref{fig: ee -> pi pi diagram}.
Figure~\ref{fig: ee -> pi pi diagram}(a)
represents the contribution from the direct $\gamma\pi\pi$
coupling $g_{\gamma\pi\pi}$.
In the present model this is expressed as
\be
g_{\gamma\pi\pi} = 1 -
\frac{Z_\rho}{2\widetilde{g}} \, g_{\rho\pi\pi}
+ y \frac{2\alpha_+}{3\widetilde{g}^2 F_{\pi p}^2} \ ,
\ee
where we also
included the term proportional to the
iso-spin violating quark mass ratio $y$.
The contribution from the $\rho$--$\omega$ mixing is shown
in fig.~\ref{fig: ee -> pi pi diagram}(c).
In section~\ref{sec: prediction},
we concentrated on the on-shell region and
used an effective momentum independent $\rho$--$\omega$ mixing.
However, here we consider the momentum dependence of the form factor,
including the momentum dependence which comes from the kinetic
type $\rho$--$\omega$ mixing provided by the $\gamma'$ term as%
\footnote{
Even if we use the momentum independent $\rho$--$\omega$ mixing,
the results in this section are not changed much
since the energy region is restricted to that near the $\rho$ and
$\omega$ mass shells.
}
\be
M_{\rho\omega}(s) = \frac{-y}{m_\rho Z_\rho^2}
\left[ 2 \alpha_+ + \nu_b - 4 \gamma' s \right]
+ \left[M_{\rho\omega}\right]_{\rm EM} \  .
\ee
Figure~\ref{fig: ee -> pi pi diagram}(d) shows the contribution from
the direct $\omega\pi\pi$ coupling $g_{\omega\pi\pi}$,
as discussed in section~\ref{sec: prediction}.

Adding the above four contributions,
the pion form factor is given by
\ba
F(s) &=&
g_{\gamma\pi\pi}
+ \frac{g_{\rho\pi\pi}}{\sqrt2} (g_\rho - C_\rho s) D_\rho(s)
- \frac{g_{\rho\pi\pi}}{\sqrt2} (g_{\omega_p} - C_{\omega_p} s)
 ( 2 m_\rho M_{\rho\omega}(s) ) D_\rho(s) D_\omega(s)
\nonum
&& {}
+ \frac{g_{\omega\pi\pi}}{\sqrt2} (g_{\omega_p} - C_{\omega_p} s)
 D_\omega(s) \ ,
\label{eq: form factor}
\ea
where
$D_V(s)$ ($V=\rho$, $\omega$) is the vector meson propagator:
\be
D_V (s) = \frac{1}{M_V^2 - s - i M_V \Gamma_V (s)} \ .
\ee
Here we use a  momentum dependent $\rho$ meson  width
but neglect this effect for the much narrower $\omega$ meson:
\be
\left\{
\begin{array}{l}
\displaystyle
\Gamma_\rho(s) =
\left(\frac{s-4m_\pi^2}{m_\rho^2-4m_\pi^2}\right)^{3/2}
\frac{m_\rho}{\sqrt{s}}\,
\theta(s-4m_\pi^2) \, \Gamma_\rho \ ,
\\
\displaystyle
\Gamma_\omega (s) = \Gamma_\omega \ .
\\
\end{array}
\right.
\ee
The parameters in the form factor
except for $C_\rho$ and $C_{\omega_p}$
are determined from the physical quantities shown in
table~\ref{tab: physical input}
together with the $\phi$--$\omega$ mixing angle $\theta_{\phi\omega}$
for fixed $x$.
[The value of $\theta_{\phi\omega}$ affects the quantities in the form
factor (\ref{eq: form factor}) through the parameter $\nu_b$.]
We determine the value of $C_{\omega_p}$ from the experimental value
of $\Gamma(\omega\rightarrow e^+e^-)$
for each $x$ as discussed in section~\ref{sec: V -> ee}.
In the previous sections,
we used the experimental value of
$\Gamma(\phi\rightarrow\pi^0\gamma)$ to determine
the $\phi$--$\omega$ mixing angle for fixed $x$ including
the $\pi^0$--$\eta$ mixing effect.
We fitted
the parameter $\nu_b$ with this angle $\theta_{\phi\omega}$.
Here, to avoid complexity and to check the
dependence on this angle,
we fix $\theta_{\phi\omega}=0.055$ or $0.0325$.
The former value is determined from the $\phi\rightarrow\pi^0\gamma$
decay width in section~\ref{sec: OZI violation} for $x$ around $20.5$,
while the latter is given in the case $\nu_b=0$ as
shown in eq.~(\ref{eq: phi omega angle}).
There are furthermore
experimental and theoretical errors in the value of the
non-electromagnetic part of the $K^+$--$K^0$ mass difference
$\left[\Delta m_K\right]_{\rm nonEM}$,
which is used to determine $y$ for fixed $x$.
We will take this uncertainty into account by using variously
$\left[\Delta m_K\right]_{\rm nonEM}=5.27$, $4.97$
and $5.57$(MeV) as inputs.

\begin{figure}[htbp]
\begin{center}
\ \epsfbox{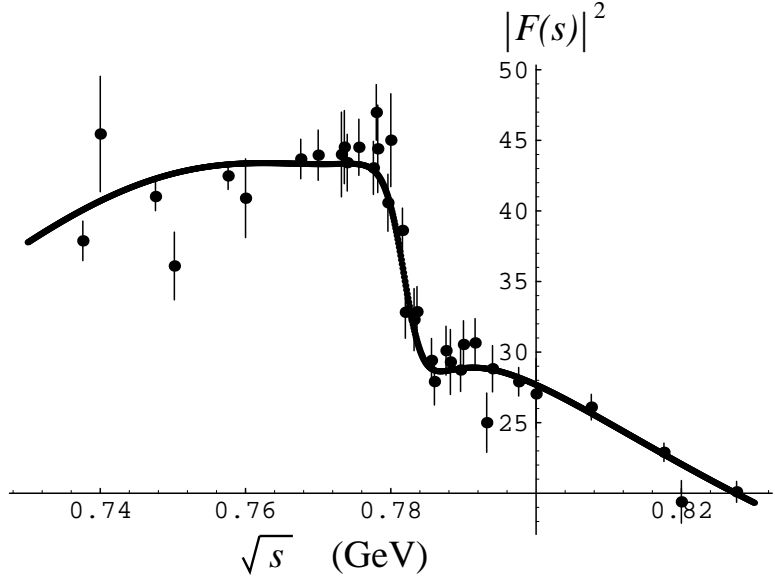}
\end{center}
\caption[]{The best fitted curve for $\theta_{\phi\omega}=0.055$
and $\left[\Delta m_K\right]_{\rm nonEM}=5.27$(MeV).
The best fitted values of $C_\rho$ and $x$ are $-0.0303$ and $17.2$,
respectively.
The experimental data is given in ref.~\cite{Barkov:85}.
}
\label{fig: eepp fit}
\end{figure}

First, we show the
best fitted curve for $\theta_{\phi\omega}=0.055$
and $\left[\Delta m_K\right]_{\rm nonEM}=5.27$(MeV)
in fig.~\ref{fig: eepp fit}.
This shows that our model fits experiment well
in the energy region where
the $\rho$ and $\omega$
mesons are close to their mass shell.

Next, we show the best fitted values of $C_\rho$ and $x$ in
table~\ref{tab: eepp fit}
for the choices of $\theta_{\phi\omega}$ and
$\left[\Delta m_K\right]_{\rm nonEM}$ mentioned above.
We also show the values of $g_{\gamma\pi\pi}$,
$C_\omega$, $R$,
$\left[M_{\rho\omega}\right]_{\rm nonEM}
\equiv M_{\rho\omega}(s\!=\!m_\rho^2) -
\left[M_{\rho\omega}\right]_{\rm EM}$
and the branching ratio
$\Gamma(\omega\rightarrow\pi\pi)/\Gamma(\rho\rightarrow\pi\pi)$.
\begin{table}[htbp]
\begin{center}
\small
\begin{tabular}{|c|c||c|c||c|c||c|c|c|}
\hline
$\theta_{\phi\omega}$
& $\left[\Delta m_K \right]_{\rm nonEM}$
& $C_\rho$ & $x$ & $g_{\gamma\pi\pi}$ & $C_\omega$
 & $R$ & $\left[M_{\rho\omega}\right]_{\rm nonEM}$
 & $\left[\Gamma(\omega)/\Gamma(\rho)\right]_{\pi\pi}$
\\
 & (MeV) & & & & & & (MeV) & $\times10^{-3}$ \\
\hline\hline
         & $4.97$ & $-0.0306$ & $20.2$ & $-0.051$ & $0.0039$
 & $51.4$ & $-2.63$ & $1.07$
\\
$0.055$  & $5.27$ & $-0.0303$ & $17.2$ & $-0.052$ & $0.0040$
 & $51.4$ & $-2.63$ & $1.07$
\\
         & $5.57$ & $-0.0300$ & $13.8$ & $-0.054$ & $0.0041$
 & $51.4$ & $-2.63$ & $1.07$
\\
\hline
         & $4.97$ & $-0.0312$ & $23.3$ & $-0.046$ & $0.0019$
 & $48.0$ & $-2.64$ & $1.06$
\\
$0.0325$ & $5.27$ & $-0.0312$ & $20.7$ & $-0.046$ & $0.0019$
 & $48.0$ & $-2.65$ & $1.06$
\\
         & $5.57$ & $-0.0311$ & $17.9$ & $-0.047$ & $0.0019$
 & $48.0$ & $-2.65$ & $1.06$
\\
\hline
\end{tabular}
\end{center}
\caption[]{The best fitted values of $C_\rho$ and $x$.
The calculated values of $g_{\gamma\pi\pi}$,
$C_\omega$,
$R$, $\left[M_{\rho\omega}\right]_{\rm nonEM}$ and
the branching ratio
$\Gamma(\omega\rightarrow\pi\pi)/\Gamma(\rho\rightarrow\pi\pi)$
are also shown.}
\label{tab: eepp fit}
\end{table}
As is expected,
the contribution from the direct $\gamma\pi\pi$ vertex is small
compared with the $\rho$ meson exchange contribution:
even in the low energy limit,
the $\rho$ meson exchange diagram gives
$g_\rho g_{\rho\pi\pi} / (\sqrt{2}m_\rho^2)\simeq1.05$,
while the direct $\gamma\pi\pi$ vertex gives
$g_{\gamma\pi\pi}\simeq-0.05$.

The best fitted value of $C_\rho$ does not depend on
$\theta_{\phi\omega}$ and
$\left[\Delta m_K\right]_{\rm nonEM}$
very much.
On the other hand,
the calculated value of $C_\omega$ does depend
on $\theta_{\phi\omega}$.
For all cases,
the absolute value of $C_\rho$ is larger than that of $C_\omega$.
In other words,
$C_\omega$ is much suppressed compared with $C_\rho$.
This implies that the $\kappa_5$ term in the kinetic type
$\rho$--$\gamma$ mixing terms (see eqs.~(\ref{eq: lag V-g })
and (\ref{fit: kinetic mixing})),
which violates the OZI rule and
breaks SU(3) symmetry at the same time,
gives a non-negligible contribution,
as discussed in section~\ref{sec: V -> ee}.

The best fitted value of $x$ depends on the choice
of input values of $\theta_{\phi\omega}$ and
$\left[\Delta m_K\right]_{\rm nonEM}$.
However, the non-electromagnetic part of the strength of
$\rho$--$\omega$ mixing
$\left[M_{\rho\omega}\right]_{\rm nonEM}$ is very stable against the
choice of input parameters:
\be
\left[M_{\rho\omega}\right]_{\rm nonEM}
= -2.63 \sim -2.65 \, \mbox{(MeV)} \, .
\ee
The estimated value of the quark mass ratio $R$ does not depend
on the choice of
$\left[\Delta m_K\right]_{\rm nonEM}$,
while it depends slightly on the $\phi$--$\omega$ mixing angle
$\theta_{\phi\omega}$.
The value of $R$ for $\theta_{\phi\omega}=0.055$ is
\be
R= 51.4 \ ,
\ee
which is larger than the best fitted value given in
section~\ref{sec: prediction}.
However, if we take the error of the
experimental value of $\omega\rightarrow\pi\pi$ decay width
into account,
these two fits agree with each other.
[The calculated value of
$\Gamma(\omega\rightarrow\pi\pi)/\Gamma(\rho\rightarrow\pi\pi)$
is $1.07\times10^{-3}$;
on the other hand, experimentally it is
$(1.23\pm0.19)\times10^{-3}$.]

Finally, we make a comment on the uncertainty of
the $\rho$ meson mass.
Since the $\rho$ meson has a broad width compared to
$\omega$ and $\phi$,
there is an uncertainty as to the precise value of its mass.
Actually the best fitted value given in ref.~\cite{Barkov:85}
is $m_\rho=775.9$(MeV).
If we fit the $\rho$ meson mass in addition to $C_\rho$ and $x$,
we obtain the following best fitted values for
$\theta_{\phi\omega}=0.055$ and
$\left[\Delta m_K\right]_{\rm nonEM}=5.27$(MeV):
\be
C_\rho = -0.0310 \ , \qquad
x = 23.1 \ , \qquad
m_\rho = 774 \, \mbox{(MeV)} \ .
\ee
These lead to $R=45.5$ and
$\Gamma(\omega\rightarrow\pi\pi)/\Gamma(\rho\rightarrow\pi\pi)
= 1.31 \times 10^{-3}$.
This $R$ agrees with the previous fit
in section~\ref{sec: prediction},
if we include the experimental error for
$\Gamma(\omega\rightarrow\pi\pi)/\Gamma(\rho\rightarrow\pi\pi)$.

This analysis shows that $x$ has an uncertainty of about $\pm3$.
Furthermore,
it indicates the close connection between the precise values of
parameters extracted from experiment and the model used to analyze the
experiment.
It is clear that obtaining greater precision in the future will
require that different experiments be analyzed with the same
theoretical model.

\newpage

\msection{Discussion}

For many purposes it is extremely useful to summarize low energy
physics up to about one GeV with an effective chiral Lagrangian.
(Evidently the pseudoscalar and vector nonet fields are the raw
materials.)
The achievement of such a goal seems to require continuous improvement
of theory (i.e., the addition of more symmetry breaking terms)
as well as the precision of the experimental results.
Here we have considered more physical processes to fit than previous
treatments and given a more complete enumeration of symmetry breaking
terms.
All parameters were determined directly from the experimental data and
a serious attempt to improve the accuracy of the analysis was made.

Naturally the treatment of symmetry breaking in the system of
pseudoscalars plus vectors is considerable more complicated than the
already complicated treatment of pseudoscalars alone.
In the latter case we have reviewed the fact that,
except for terms needed\footnote{
These terms are discussed
in ref.~\cite{Schechter-Subbaraman-Weigel}.
}
to solve the U(1) problem,
the chiral perturbation theory
analysis\cite{Weinberg:ChPT,Gasser-Leutwyler:SU(2)-SU(3)}
of chiral symmetry breaking is essentially reproduced at tree level
with three OZI rule conserving terms of type
${\cal M}$, ${\cal M}\partial^2$ and ${\cal M}^2$.
In practice it seems extremely difficult to definitely establish the
existence of non-negligible loop corrections at the level of the one
and two point functions involved in the analysis of symmetry breaking
(One must go to $\pi\pi$ scattering for this purpose).
Such a situation might be expected if the $1/N_c$ approximation is
quite good.
We have given an analogous treatment of the pseudoscalar plus vector
system, including only tree diagrams but allowing higher order
symmetry breaking terms which may mask any loop effects.
We did, however, give some discussion and speculation on how to
provide a controllable expansion scheme in this more complicated
situation.
We also showed (section~\ref{sec: V -> ee}) how the prediction for the
decay $\rho^0\rightarrow e^+e^-$ could be significantly improved
either with higher derivative terms or by an old
calculation\cite{Gounaris-Sakurai}
which amounts to the partial inclusion of the loop effects.
It would be interesting to further investigate this
and related processes in the
future.

An interesting question is whether the symmetry breaking patterns
(modulo the important but understood difference in the strengths of
OZI violation)
for the vectors and pseudoscalars are precisely analogous.
In other words,
do we really need to include ``second order'' symmetry breaker for the
vector nonet?
Here we have seen indications that they are needed.
Apart from the explanation of the $\rho^0\rightarrow e^+e^-$
rate mentioned above,
the ${\cal M}\partial^2$ type $\gamma'$ term in
eq.~(\ref{Lag: first}) was seen to be helpful for improving the
predicted $K^0$ ``charge radius''.
The jury
is still out on the need for ${\cal M}^2$ type vector symmetry
breakers.
We saw that they could increase the size of the
$K^{\ast0}$--$K^{\ast+}$ mass difference
$\Delta m_{K^{\ast}}$ by a small amount.
This would move the prediction in the direction towards the central
experimental values.
As discussed in detail in section~\ref{sec: prediction}
it is possible that experiment and theory are already in agreement for
$\Delta m_{K^{\ast}}$
but further theoretical analysis of the electomagnetic contribution
and a more refined experimental analysis seems necessary.

It should be remarked that the symmetry breaking terms
(even though slightly numerous) do not
{\em drastically} \/ change the
simple picture,
present in the model without symmetry breaking, of
vector meson dominance and the associated empirical KSRF formula.
Rather they provide ``fine tuning''.
There are just a few {\em essential} \/ parameters
needed to obtain an approximate fit.
Furthermore, for  the fit presented here,
rather conventional values (see
eqs.~(\ref{Rval 1}) and (\ref{quark mass ratio}))
of the quark mass ratios were extracted.
The quark mass ratio determination
is only subject to the uncertainty of the
``$\Delta m_{K^{\ast}}$ puzzle''
(sections~\ref{sec: prediction} and \ref{sec: second order}).
It is interesting to notice that ,
for the small vector nonet OZI rule violation piece of the
Lagrangian,
both SU(3) symmetry breaking as well as SU(3) symmetric terms are
required.
There does not seem to be a multiplicative suppression of OZI
violating $\times$ SU(3) violating terms.

We have not presented in this paper an analysis of symmetry breaking
effects for the terms of the Lagrangian proportional to the
Levi-Civita symbol.
This topic will be discussed elsewhere.

\vspace{0.5cm}

\section*{Acknowledgements}

We would like to thank Francesco Sannino,
Anand Subbaraman and Herbert Weigel for helpful discussions.
This work was supported in part by the U.S. DOE Contract
No. DE-FG-02-85ER40231.
One of us (M.H.) acknowledges partial
support by the Japan Society for the
Promotion of Science.

\newpage

\appendix

\msection{Formulae
\label{app: formula}}

In this appendix, we list the quantities discussed in this paper
computed
from our Lagrangian:
$\Lag = \Lag_{\rm sym} + \Lag_{\rm SB} +
\Lag_{\nu1} + \Lag_{\nu2}$.
In the appropriate limits the formulae reduce to those given in
ref.~\cite{Schechter-Subbaraman-Weigel};
one should replace:
$\alpha_+\rightarrow\alpha'$, $\alpha_-\rightarrow\alpha'$
and $\alpha_p\rightarrow\alpha'-4\beta'\widetilde{g}^2$.

\subsection{Alternate form of Lagrangian}
\label{app: rewriting}

Here for comparison with other papers
we rewrite our Lagrangian by using the following fields:
\ba
p_\mu &\equiv& \frac{i}{2}
 \left( \xi \delm \xi^{\dag} - \xi^{\dag} \delm \xi \right) \ ,
\nonum
v_\mu &\equiv& \frac{i}{2}
 \left( \xi \delm \xi^{\dag} + \xi^{\dag} \delm \xi \right) \ .
\ea
For the symmetry breaking spurion, the following combinations are
convenient:
\be
\widehat{\cal M}_{\pm}
\equiv \frac{1}{2} \left(
  \xi \massd \xi \pm \xi^{\dag} \mass \xi^{\dag} \right) \ .
\label{def: mass p m}
\ee
Using the quantities $v_\mu$ and $p_\mu$, the
symmetric Lagrangian (\ref{Lag: sym}) is rewritten as
\be
\Lag_{\rm sym} =
- \frac{1}{2} m_v^2
\Tr\left[ \widehat{\rho}_\mu \, \widehat{\rho}_\mu \right]
- \frac{F_\pi^2}{2} \Tr\left[ p_\mu \, p_\mu \right]
- \frac{1}{4}
\Tr \left[ F_{\mu\nu}(\rho) F_{\mu\nu}(\rho) \right]
\ ,
\ee
where
$\widehat{\rho}_\mu \equiv \rho_\mu - v_\mu / \gtil$.
The $\alpha_i$ terms in the
first order symmetry breaking terms are rewritten as
\be
\Lag_{\alpha_i} =
2 \alpha_+
\Tr\left[ \mhatp
  \widehat{\rho}_\mu \, \widehat{\rho}_\mu
\right]
- 2 \frac{\alpha_-}{\gtil}
\Tr\left[ \mhatm
  \left[\widehat{\rho}_\mu \, , \,p_\mu \right]
\right]
-2  \frac{\alpha_p}{\gtil^2}
\Tr\left[ \mhatp p_\mu p_\mu \right] \ ,
\ee
where $\alpha_+$, $\alpha_-$ and $\alpha_p$ are defined in
table~\ref{tab: new para}.
The $\mu$ terms in eq.~(\ref{Lag: second})
are rewritten as
\ba
\Lag_{\mu} &=&
\mu_a \Tr \left[
  \widehat{\rho}_\mu \mhatp \widehat{\rho}_\mu \mhatp
\right]
-
\mu_b \Tr \left[
  \widehat{\rho}_\mu \mhatm \widehat{\rho}_\mu \mhatm
\right]
-
\frac{\mu_c}{\gtil^2} \Tr \left[
  p_\mu \mhatp p_\mu \mhatp
\right]
\nonum
&& +
\frac{\mu_d}{\gtil^2} \Tr \left[
  p_\mu \mhatm p_\mu \mhatm
\right]
-
\frac{2\mu_e}{\gtil} \Tr \left[
  \widehat{\rho}_\mu \mhatp p_\mu \mhatm
  - \widehat{\rho}_\mu \mhatm p_\mu \mhatp
\right] \ ,
\label{eq: mu terms 2}
\ea
where
$\mu_a$, $\mu_b$, $\mu_c$, $\mu_d$ and $\mu_e$ are defined in
table~\ref{tab: new para}.
The OZI rule violating terms
in eqs.~(\ref{Lag: nu 1}) and (\ref{Lag: nu 2})
are also rewritten as
\be
\Lag_{\nu} =
\nu_a
\left( \Tr\left[\widehat{\rho}_\mu \right] \right)^2
+\nu_b
\Tr\left[\widehat{\rho}_\mu \right]
\Tr\left[\mhatp \widehat{\rho}_\mu \right]
+ \frac{\nu_c}{\gtil^2}
\left( \Tr\left[ p_\mu \right] \right)^2
+ \frac{\nu_d}{\gtil^2}
\Tr[p_\mu] \Tr\left[ \mhatp p_\mu \right] \ ,\
\label{Lag: nu term}
\ee
where $\nu_a$, $\nu_b$, $\nu_c$ and $\nu_d$
are defined in table~\ref{tab: new para}.
Note that only the $\nu_a$ and $\nu_b$ terms are relevant for the
vector nonet.

\subsection{kinetic terms}

The wave function renormalization constants
for pseudoscalar and vector mesons
defined in eq.~(\ref{eq: field renorm})
are given by
\be
Z_\pi =
\left[
  1 + \frac{2 (2\alpp+\mu_c)}{\gtil^2 F_\pi^2}
\right]^{1/2} \ , \
Z_K =
\left[
  1 + \frac{2 (1+x) \alpp + 2x\mu_c }{\gtil^2 F_\pi^2}
\right]^{1/2} \ .
\label{def: Zpi}
\ee
and
\be
Z_\rho = Z_\omega = \left[ 1- 8 \gamma' \right]^{1/2} \ ,\
Z_\phi = \left[ 1- 8 x \gamma' \right]^{1/2} \ , \
Z_{K^{\ast}} = \left[ 1- 4 \gamma' (1+x) \right]^{1/2} \ .
\label{def: Zrho}
\ee

\subsection{masses and mixings}

The vector meson masses are given by
\ba
&& m_\rho^2 =
\left[m_v^2-4\alpha_+ -2\mu_a\right]/Z_\rho^2 \ ,
\nonum
&& m_\omega^2 =
\left[m_v^2-4\alpha_+ -2\mu_a - 4(\nu_a+\nu_b)\right]/Z_\omega^2 \ ,
\nonum
&& m_\phi^2 =
\left[m_v^2-4x\alpha_+ -2x^2\mu_a-2(\nu_a+x\nu_b)\right]/Z_\phi^2 \ ,
\nonum
&& m_{K^{\ast}}^2 =
\left[ m_v^2 - 2(1+x)\alpha_+ - 2x\mu_a \right]/Z_{K^{\ast}}^2 \ .
\label{eq: vector masses}
\ea
The non-electromagnetic $K^0$--$K^+$ and
$K^{0\ast}$--$K^{+\ast}$ mass differences
are given by
\ba
&&
\left[\Delta m_K\right]_{\rm nonEM} \equiv
\left[ m_{K^0} - m_{K^+} \right]_{\rm nonEM}
\nonum
&& \mq =
\frac{y}{F_{K p}^2 m_K}
\Biggl[
  - 4 \delta' - 16 (1+x) {\lambda'}^2
  + 2 m_K^2 \frac{\alpp + x\mu_c}{\gtil^2}
\Biggr] \ . \
\label{eq: K0 - K+ mass diff}
\\
&&
\left[\Delta m_{K^{\ast}}\right]_{\rm nonEM} \equiv
\left[m_K^{0\ast}-m_K^{+\ast}\right]_{\rm nonEM} =
y \left[ 2\alpha_+ - 4 \gamma' m_{K^{\ast}}^2 + 2x\mu_a \right]
/ (m_{K^{\ast}} Z_{K^{\ast}}^2)\ .
\label{eq: K star mass diff}
\ea
The $\rho^0$--$\omega$ transition mass $M_{\rho\omega}$
is defined in
terms of the effective term in the Lagrangian:
$-2 m_\rho M_{\rho\omega}\rho^0_\mu\omega_\mu$.
This is given by
\be
\left[M_{\rho\omega}\right]_{\rm nonEM} =
- y \left[
  2\alpha_+ - 4\gamma' m_\rho^2 + 2 \mu_a + \nu_b
\right]/ (m_\rho Z_\rho^2)\ .
\label{eq: rho omega mix}
\ee
Similarly, $\phi$--$\omega$ and $\phi$--$\rho$ mixings,
which are defined by $\Lag=-\Pi_{\phi\omega}\phi_\mu\omega_\mu$
and $\Lag=-\Pi_{\phi\rho}\phi_\mu\rho_\mu$,
are given by
\ba
\Pi_{\phi\omega} &=&
\frac{-\sqrt{2}\left(2\nu_a+(1+x)\nu_b\right)}{Z_\omega Z_\phi} \ ,
\label{eq: phi omega mix}\\
\Pi_{\phi\rho} &=&
\frac{-\sqrt{2}y\nu_b}{Z_\rho Z_\phi} \ .
\label{eq: phi rho mix}
\ea

\subsection{Vector-Pseudoscalar couplings}

We define the $VPP'$ coupling constant
$g_{\scriptscriptstyle V P P'}$ by
\be
\Lag = - \frac{i}{\sqrt{2}} g_{\scriptscriptstyle V P P'}
V_\mu \left( \delm P P' - \delm P' P \right) \ .
\label{def: V PP}
\ee
We list the forms of these couplings:
\ba
&& g_{\rho^0\pi^+\pi^-} =
\Bigl[
  m_v^2 - 4 \alpha_+ + 8 \alpha_- - 2\mu_a - 8 \mu_e
\Bigr] /(\gtil F_{\pi p}^2 Z_\rho)\ ,
\nonum
&& g_{\rho^0{\scriptscriptstyle K^+ K^-}} = \frac{1}{2}
\Bigl[
  m_v^2 - 4\alpha_+ + 4(1+x)\alpha_- - 2\mu_a
  - 4(1+x)\mu_e
\Bigr] /(\gtil F_{Kp}^2 Z_\rho) \ ,
\nonum
&& g_{\rho^0 {\scriptscriptstyle K^0 \overline{K}^0}}
= - g'_{\rho^0 {\scriptscriptstyle K^+ K^-}} \ ,
\nonum
&& g_{\omega {\scriptscriptstyle K^+ K^-}} = \frac{1}{2}
\Bigl[
  m_v^2 - 4\alpha_+ + 4(1+x)\alpha_- - 2\mu_a
  - 4(1+x)\mu_e + 2(x-1) \nu_b
\Bigr] /(\gtil F_{Kp}^2 Z_\rho) \ ,
\nonum
&& g_{\omega {\scriptscriptstyle K^0 \overline{K}^0}}
= g_{\omega {\scriptscriptstyle K^+ K^-}} \ ,
\nonum
&& g_{\phi {\scriptscriptstyle K^+ K^-}} = - \frac{1}{\sqrt{2}}
\Bigl[
  m_v^2 - 4x \alpha_+ + 4(1+x) \alpha_- - 2x^2\mu_a
\nonum && \mqqq\mqqq
  - 4x(1+x)\mu_e - (x-1)\nu_b
\Bigr] /(\gtil F_{Kp}^2 Z_\phi) \ ,
\nonum
&& g_{\phi {\scriptscriptstyle K^0 \overline{K}^0}} =
g_{\phi {\scriptscriptstyle K^+ K^-}} \ ,
\nonum
&& g_{{\scriptscriptstyle K^{\ast+}K^-}\pi^0} = \frac{1}{2}
\Bigl[
  m_v^2 - 2(1+x)\alpha_+ + 2(x+3)\alpha_- - 2x\mu_a
  - 2(3x+1)\mu_e
\Bigr] /(\gtil F_{Kp}F_{\pi p} Z_{K^{\ast}}) \ ,
\nonum
&& g_{{\scriptscriptstyle K^{\ast+}\overline{K}^0}\pi^-} =
g_{{\scriptscriptstyle K ^{\ast0}K^-}\pi^+} =
\sqrt{2} g_{{\scriptscriptstyle K^{\ast+}K^-}\pi^0}  \ ,
\nonum
&& g_{{\scriptscriptstyle K^{\ast0}\overline{K}^0}\pi^0} =
g_{{\scriptscriptstyle K^{\ast+}K^-}\pi^0} \ .
\label{eq: V phi phi coupling}
\ea
Here for later convenience we define the following
$VPP'$ couplings $g_{\scriptscriptstyle VPP'}$
\be
g_{\phi {\scriptscriptstyle K\overline{K}}} =
- \sqrt{2} g_{\phi {\scriptscriptstyle K^+ K^-}}\, , \mq
g_{{\scriptscriptstyle K^{\ast}K}\pi} =
2 g_{{\scriptscriptstyle K^{\ast+}K^-}\pi^0}\ .
\ee
The isospin breaking vertices are given by
\ba
&& g_{\omega\pi\pi} = - 2y
\Bigl[
  2 \alpha_+ + 2\mu_a + 4 \mu_e + \nu_b
\Bigr] /(\gtil F_{\pi p}^2 Z_\omega)\ ,
\label{def: direct omega pipi}
\\
&& g_{\phi\pi^+\pi^-} = - y
\frac{\sqrt{2} \nu_b}{\gtil F_{\pi p}^2 Z_\phi} \ .
\label{def: direct phi pipi}
\ea

\subsection{$V$--$\gamma$ transition terms}

The expressions for
the vector meson--photon
transition strengths
defined in eq.~(\ref{eq: V - gamma mixing})
are given by
\ba
g_\rho &=&
\frac{1}{Z_\rho} \frac{m_v^2-4\alpha_+-2\mu_a}{\sqrt{2}\gtil}
= \frac{m_\rho^2Z_\rho}{\sqrt{2}\gtil}
\ ,
\nonum
g_\omega &=&
\frac{1}{3} \frac{1}{Z_\omega}
\frac{m_v^2-4\alpha_+-2\mu_a+2(x-1)\nu_b}{\sqrt{2}\gtil}
= \frac{1}{3} \frac{m_\omega^2Z_\rho}{\sqrt{2}\gtil}
\left[
  1 - \frac{Z_\phi}{Z_\rho}
  \frac{\sqrt{2}\Pi_{\phi\omega}}{m_\omega^2}
\right]
\ ,
\nonum
g_\phi &=&
- \frac{\sqrt2}{3} \frac{1}{Z_\phi}
\frac{m_v^2-4x\alpha_+-2x^2\mu_a-(x-1)\nu_b}{\sqrt{2}\gtil}
= - \frac{\sqrt2}{3} \frac{m_\phi^2Z_\phi}{\sqrt{2}\gtil}
\left[
   1 - \frac{Z_\rho}{Z_\phi}
   \frac{\sqrt{2}\Pi_{\phi\omega}}{2m_\phi^2}
\right]
\ . \ \ \ \ \ \
\label{eq: mixing strengths}
\ea

\newpage

\msection{Details of $\phi\rightarrow\pi\gamma$ calculation
\label{app: phi pi gamma}}

This process is important for estimating the strength of the
$\phi_\mu$--$\omega_\mu$ mixing coefficient $\Pi_{\phi\omega}$.
As a potentially non-negligible correction we also compute the
$\pi^0$--$\eta$ mixing mediated contribution shown in
fig.~\ref{fig: phi -> pi gamma}(b).
In ref.~\cite{Jain-Johnson-Meissner-Park-Schechter}
the $\phi$--$\omega$ mixing was considered to be the only source for
this decay.

\subsection{$\pi^0$--$\eta$ mixing}

First we describe
the $\pi^0$--$\eta$ mixing.
It is convenient to define
$\eta_{\scriptscriptstyle T} \equiv
(\phi_{11} + \phi_{22})/\sqrt{2}$ and
$\eta_{\scriptscriptstyle S} \equiv \phi_{33}$.
Following ref.~\cite{Schechter-Subbaraman-Weigel},
let us write the
relation between these fields and the physical $\eta$ and $\eta'$
fields as
\ba
&& \!\!\!\!\!\!\!\!\!\!\!\!
\left( \begin{array}{c}
 \eta_{\scriptscriptstyle T} \\ \eta_{\scriptscriptstyle S} \\
\end{array} \right)
=
\left(
\begin{array}{cc}
 A_{11} & A_{12} \\
 A_{21} & A_{22} \\
\end{array}
\right)
\left(
\begin{array}{c}
 \eta \\
 \eta' \\
\end{array}
\right) \ ,
\nonumber\\
&& \!\!\!\!\!\!\!\!\!\!
\left(
\begin{array}{cc}
 A_{11} & A_{12} \\
 A_{21} & A_{22} \\
\end{array}
\right)
=
\left(
\begin{array}{cc}
 \cos \theta_1 & \sin \theta_1 \\
 - \sin \theta_1 & \cos \theta_1 \\
\end{array}
\right)
\left(
\begin{array}{cc}
 \widehat{K}^{-1/2}_1 & 0 \\
 0 & \widehat{K}^{-1/2}_2 \\
\end{array}
\right)
\left(
\begin{array}{cc}
 \cos \theta_2 & \sin \theta_2 \\
 - \sin \theta_2 & \cos \theta_2 \\
\end{array}
\right)  \ .\
\label{def: eta eta' mixing}
\ea
In this paper we shall use the values
for these parameters shown in
ref.~\cite{Schechter-Subbaraman-Weigel}:
\be
\theta_1 = 7.44^\circ \ , \quad
\theta_2 = 34.7^\circ \ , \quad
\widehat{K}_1^{1/2} = 1.07 \ , \quad
\widehat{K}_2^{1/2} = 1.36 \ .
\label{value: eta eta' mixing}
\ee
In the present model, there are both mass type and kinetic type
$\pi^0$--$\eta_{\scriptscriptstyle T}$ mixings.
These are given by
\be
\Lag =
- K_{\pi\eta} \partial_\mu \pi^0
  \partial_\mu \eta_{\scriptscriptstyle T}
- \Pi_{\pi\eta} \pi^0 \eta_{\scriptscriptstyle T} \ ,
\ee
where
\be
K_{\pi\eta} = 4 y \frac{\alpha_p}{\gtil^2F_\pi F_{\pi p}} \ ,
\qquad
\Pi_{\pi\eta} = \frac{8y}{F_\pi F_{\pi p}}
\left[ \delta' + 8 {\lambda'}^2 \right] \ .
\label{def: eta pi mixing}
\ee
After diagonalizing the $\eta$--$\eta'$ part as shown in
eq.~(\ref{def: eta eta' mixing}),
we include the $\pi^0$--$\eta$ mixing effect by introducing
the following physical fields:
\be
\left( \begin{array}{c}
 \pi^0 \\ \eta\\
\end{array} \right)
=
\left(
\begin{array}{cc}
 1 & \varepsilon_{12} \\ \varepsilon_{21} & 1 \\
\end{array}
\right)
\left(
\begin{array}{c}
 \pi^0_p \\ \eta_p \\
\end{array}
\right) \ ,
\ee
where
\be
\varepsilon_{12} = A_{11}
\frac{\Pi_{\pi\eta} - K_{\pi\eta} m_\eta^2}%
{m_\eta^2 - m_\pi^2} \ ,
\qquad
\varepsilon_{21} = - A_{11}
\frac{\Pi_{\pi\eta} - K_{\pi\eta} m_\pi^2}%
{m_\eta^2 - m_\pi^2} \ .
\label{eq: eta pi mixings}
\ee

\subsection{$\phi\rightarrow\pi\gamma$ decay}

The partial width of the
$V\rightarrow P \gamma$ decay process
($V$: vector meson, $P$: pseudoscalar meson)
is given by
\be
\Gamma(V\rightarrow P \gamma)
= \frac{3\alpha \vert q_\gamma(V)\vert^3}{32\pi^4 F_{\pi p}^2}
\left\vert G_{VP\gamma} \right\vert^2 \ ,
\ee
where $q_\gamma(V) = (m_V^2-m_P^2)/(2m_V)$
and
$G_{VP\gamma}$
is defined by
\be
\Lag =
\frac{3 e}{4\sqrt{2}\pi^2F_{\pi p}}
G_{VP\gamma} \, \varepsilon_{\mu\nu\lambda\sigma}\,
\partial_\mu V_\nu \,
\partial_\lambda {\cal A}_\sigma \, P \ .
\ee
Using the experimental values
$\Gamma(\omega\rightarrow\pi^0\gamma) = 0.72$(MeV) and
$\Gamma(\phi\rightarrow\eta\gamma) = 5.8$(KeV),
we find
$
\left\vert G_{\omega\pi\gamma} \right\vert
= 5.7$,
$\left\vert G_{\phi\eta\gamma} \right\vert
= 1.7$.
If we consider that $\phi$--$\omega$ mixing and
$\pi$--$\eta$ mixing are the sources for
$\phi\rightarrow\pi^0\gamma$ decay,
the effective coupling for this decay is related to
these two couplings as
\be
G_{\phi\pi\gamma}
 = \frac{\Pi_{\phi\omega}}{m_\phi^2-m_\omega^2}
G_{\omega\pi\gamma}
+ \varepsilon_{21} G_{\phi\eta\gamma} \ ,
\ee
where
$\Pi_{\phi\omega}$ is given
in eq.~(\ref{eq: phi omega mix})
while $\varepsilon_{21}$ is given in eq.~(\ref{eq: eta pi mixings}).

To determine the value of
$G_{\phi\pi\gamma}$
we need to know
the relative sign between
$G_{\omega\pi\gamma}$ and
$G_{\phi\eta\gamma}$.
As shown,
for example, in refs.~%
\cite{Fujiwara-Kugo-Terao-Uehara-Yamawaki,%
Jain-Johnson-Meissner-Park-Schechter}
both terms are part of a single trace and one obtains
$R_{\phi\omega} \equiv
G_{\phi\eta\gamma}/G_{\omega\pi\gamma}
=
- 2 A_{21}/3$,
where $A_{21}$ is a component of the $\eta$--$\eta'$ mixing
matrix.
Using the values in eqs.~(\ref{value: eta eta' mixing})
we find $A_{21}=-0.52$, and
$R_{\phi\omega}=0.35$
which agrees with the experiment
(Experimentally $\vert R_{\phi\omega} \vert = 0.30$).
Then we can take both
$G_{\omega\pi\gamma}$ and
$G_{\phi\eta\gamma}$ to be positive for present purposes and set
\be
G_{\omega\pi\gamma}
= 5.7 \ , \qquad
G_{\phi\eta\gamma}
= 1.7 \ ,
\qquad
A_{11} = 0.71 \ ,
\ee
where for $A_{11}$ we use
the values in eqs.~(\ref{value: eta eta' mixing}).

\newpage

\msection{$\omega\rightarrow\pi^+\pi^-$ and
$\phi\rightarrow\pi^+\pi^-$
decay processes
\label{app: omega pi pi}}

There are two main contributions to the isospin breaking decay
$\omega\rightarrow\pi^+\pi^-$:
(1)
$\rho$--$\omega$ mixing $M_{\rho\omega}$;
(2)
the isospin breaking direct $\omega\pi^+\pi^-$ vertex
$g_{\omega\pi\pi}$.
As discussed in ref.~\cite{Gasser-Leutwyler:PRep},
the masses of the
$\rho$ and $\omega$ mesons are so close that it is
important to take the widths into account for the effect of
$\rho$--$\omega$ mixing.
Then
the ratio of $\pi\pi$ decay width of $\omega$ to
that of $\rho$ is given by
\ba
&&
\frac{\Gamma(\omega\rightarrow\pi^+\pi^-)}%
{\Gamma(\rho\rightarrow\pi^+\pi^-)}
=
\left\vert
 \frac{g_{\omega\pi\pi}}{g_{\rho\pi\pi}}
 + \frac{M_{\rho\omega}}%
 {m_\omega - m_\rho - \frac{i}{2}(\Gamma_\omega - \Gamma_\rho)}
\right\vert^2
\abs{\frac{q_\pi(\omega)}{q_\pi(\rho)}}^3
\frac{m_\rho^2}{m_\omega^2}
\ ,
\label{eq: rho omega}
\ea
where
$q_\pi(V) \equiv \frac{1}{2}\sqrt{ m_V^2 - 4 m_\pi^2 }$,
($V=\rho$, $\omega$).
The experimental value of this ratio is estimated as
$(1.23 \pm 0.19) \times 10^{-3}$.
For the experimental fit of this process
it is important to include
the electromagnetic contribution to the $\rho$--$\omega$
mixing.
This is estimated as\cite{Gasser-Leutwyler:PRep}
\ba
\left[M_{\rho\omega}\right]_{\rm EM} &= &
\frac{3}{2\alpha} \sqrt{\frac{m_\omega}{m_\rho}}
\sqrt{\Gamma(\rho\rightarrow e^+e^-)\,%
\Gamma(\omega\rightarrow e^+e^-)}
\simeq
(0.417\pm0.017) \,\mbox{\rm MeV} \ .
\label{eq: rho - omega from photon}
\ea
For comparison, we consider the case where
the direct $\omega\pi\pi$ vertex is negligible
($g_{\omega\pi\pi}\simeq0$).
Using the above formula (\ref{eq: rho omega}) with
$g_{\omega\pi\pi}=0$, we obtain
\be
M_{\rho\omega} = - (2.50\pm0.20) \,\mbox{\rm MeV} \ , \qquad
\left[ M_{\rho\omega} \right]_{\rm nonEM}
= - (2.92\pm0.22) \,\mbox{\rm MeV} \ .
\ee

Similarly, for the $\phi\rightarrow\pi^+\pi^-$
process, the
main contribution is given by the $\phi\rho$ mixing
$\Pi_{\phi\rho}$
and there is a relatively small
direct $\phi\pi\pi$ coupling $g_{\phi\pi\pi}$.
We then have
\be
\frac{\Gamma(\phi\rightarrow\pi^+\pi^-)}%
{\Gamma(\rho\rightarrow\pi^+\pi^-)}
=
\left\vert
 \frac{g_{\phi\pi\pi}}{g_{\rho\pi\pi}} +
 \frac{\Pi_{\phi\rho}}{m_\phi^2-m_\rho^2}
\right\vert^2
\abs{\frac{\bfq_\pi(\phi)}{\bfq_\pi(\rho)}}^3
\frac{m_\rho^2}{m_\phi^2} \ .
\label{eq: phi pi pi decay}
\ee
Experimentally, this ratio is estimated as
$\left(2.3^{+1.5}_{-1.2}\right)\times 10^{-6}$.
The electromagnetic contribution to the $\phi$--$\omega$ mixing is
estimated as
\be
\left[\Pi_{\phi\omega}\right]_{\rm EM}
= \frac{3}{\alpha} m_\omega \frac{m_\omega}{m_\phi}
\sqrt{ \Gamma(\phi\rightarrow e^+e^-)
  \Gamma(\omega\rightarrow e^+e^-)
}
\simeq
(0.255\pm0.006) \times 10^{-3} \ (\mbox{GeV}^2) \ .
\ee
This is very small compared with the nonelectromagnetic contribution,
and we neglect it in this paper.

\end{document}